\documentclass[sigconf]{acmart}
\AtBeginDocument{%
  }

\usepackage{graphicx}
\usepackage{amsmath}
\usepackage{booktabs}
\usepackage{multirow}
\usepackage{enumitem}
\usepackage{subcaption}
\usepackage{colortbl}
\usepackage[most]{tcolorbox}
\usepackage{xcolor}
\usepackage{tabularx}
\usepackage{siunitx}
\usepackage{hyperref}

\definecolor{effectLarge}{RGB}{255,135,131}     
\definecolor{effectMedium}{RGB}{255,187,84}   
\definecolor{effectSmall}{RGB}{255,243,125}    

\newtcolorbox{rqsummary}{
    colback=gray!5,
    colframe=gray!40,
    fonttitle=\bfseries,
    coltitle=black,
    fontupper=\scriptsize
}

\usepackage{pifont}      

\definecolor{pos}{HTML}{E8F5E9}   
\definecolor{neg}{HTML}{FFEBEE}   
\definecolor{neu}{HTML}{F7F7F7}   
\definecolor{sig}{HTML}{FFF8E1}   

\newcommand{\up}{\textcolor{green!55!black}{\ding{115}}}   
\newcommand{\down}{\textcolor{red!70!black}{\ding{116}}}   
\newcommand{\neutraldir}{\textcolor{gray!70}{\textbullet}}

\newcolumntype{L}{>{\raggedright\arraybackslash}X}
\newcolumntype{R}{>{\raggedleft\arraybackslash}X}
\newcolumntype{D}{>{\centering\arraybackslash}p{0.2cm}}

\newcommand{\swebenchverified}{SWE-bench Verified}
\newcommand{\multiswebench}{Multi-SWE-bench}
\begin{document}

\title{How Do LLMs Read Bug Reports? An Empirical Study of Attention in LLMs for Automated Program Repair}


\author{Ramtin Ehsani}
\email{ramtin.ehsani@drexel.edu}
\orcid{0000-0003-1517-7135}
\affiliation{%
  \institution{Drexel University}
  \city{Philadelphia}
  \state{PA}
  \country{USA}
}

\author{Irene Manotas}
\email{irene.manotas@ibm.com}
\affiliation{%
  \institution{IBM Research}
  \city{Yorktown}
  \state{NY}
  \country{USA}}

\author{Saurabh Pujar}
\email{saurabh.pujar@ibm.com}
\affiliation{%
  \institution{IBM Research}
  \city{Yorktown}
  \state{NY}
  \country{USA}}

\author{Luca Buratti}
\email{luca.buratti1@ibm.com}
\affiliation{%
  \institution{IBM Research}
  \city{Yorktown}
  \state{NY}
  \country{USA}}

\author{Preetha Chatterjee}
\email{preetha.chatterjee@drexel.edu}
\affiliation{%
  \institution{Drexel University}
  \city{Philadelphia}
  \state{PA}
  \country{USA}
}
\renewcommand{\shortauthors}{Ehsani et al.}

\begin{abstract}
Large Language Model (LLM)-based Automated Program Repair systems are advancing rapidly, yet their performance remains inconsistent. Even when provided with the same contextual information, an LLM may generate a correct patch for one bug but fail on another closely related bug. Why this happens remains poorly understood, and it is unclear how LLMs prioritize the diverse information in bug reports and whether model attention affects repair success.

In this paper, we present the first empirical study of attention patterns in LLM-based program repair, providing interpretable insights into how models process bug reports and where their attention is concentrated during repair.
We analyze 319 real-world Python and Java bugs from \textit{\swebenchverified{}} and \textit{\multiswebench{}} to study \textbf{(RQ1)} how model attention is distributed across bug report sections, \textbf{(RQ2)} how attention patterns within each section differ between successful and unsuccessful repairs, and \textbf{(RQ3)} how these patterns compare to information developers consider important for bug fixing.
We find that successful repairs are characterized by diffused attention across multiple diagnostic components such as bug descriptions, stacktraces, and test cases, while failures often exhibit over-localized attention toward metadata such as version information. We further observe that stronger alignment between model attention and developer-identified key sections and phrases is associated with higher repair success.
Our results provide the first empirical evidence that attention misallocation is a key factor in LLM-based APR failures, and offer actionable insights for designing more interpretable and reliable future APR systems.
\end{abstract}

\ccsdesc[500]{Software and its engineering~Software maintenance tools}
\ccsdesc[500]{Software and its engineering~Automatic programming}
\ccsdesc[500]{Computing methodologies~Artificial intelligence}

\keywords{large language models, attention analysis, program repair}


\maketitle

\section{Introduction}
LLMs are now deeply embedded in software development tasks such as Automated Program Repair (APR), which involves automatically generating code patches to fix software bugs~\cite{10.1145/3764584}. Given a bug report and the corresponding buggy code, LLMs attempt to produce a patch that resolves the bug while preserving the intended program behavior.
Although LLM-based program repair has shown significant improvements over traditional techniques \cite{ehsani2025hierarchicalknowledgeinjectionimproving, xia2025livesweagentsoftwareengineeringagents, 10.1109/ICSE48619.2023.00129}, failure cases remain frequent and unpredictable~\cite{liang-etal-2025-language}. A model may successfully fix one bug while failing on a nearly identical one, providing no insight into its decision-making process~\cite{li2025evaluatinggeneralizabilityllmsautomated, ehsani2025hierarchicalknowledgeinjectionimproving}. This lack of interpretability limits our ability to understand and integrate LLM-based repair systems into developer workflows.

Bug reports contain multiple sections, such as a bug description, steps to reproduce the failure, expected behavior, and other information. 
Repairing real bugs is a complex task, particularly for LLMs, which must attend to the right type and amount of information to succeed~\cite{ehsani2025hierarchicalknowledgeinjectionimproving}. To generate a correct patch, models must infer the underlying cause of failure, reason about the intended behavior, and integrate signals distributed across multiple sections of the bug report. 
Recent studies suggest that errors in code generation tasks often stem from how LLMs allocate attention to different types of input information~\cite{kou2024large, zhang2024tellmodelattendposthoc, 10.1145/3696630.3728510, 11334352}. This raises the question of whether a similar phenomenon also explains failures in LLM-based program repair. 
While prior studies have analyzed model attention for code generation from short task descriptions~\cite{kou2024large, li2024machines, 10645745, ning2024insights}, extending such analysis to real-world bug reports is substantially more challenging because bug reports are often long, heterogeneous, and contain both code and natural-language information. 
Understanding where models focus their attention can reveal which parts of a bug report most influence repair decisions.


In this paper, we present the first empirical study of how LLMs allocate attention to bug reports during automated program repair. We analyze 319 Python and Java bugs \textcolor{black}{with different levels of difficulty (\textit{Easy}, \textit{Medium}, and \textit{Hard})} from \textit{\swebenchverified} and \textit{\multiswebench} using both proprietary (claude-4-sonnet~\cite{anthropic}) and open-source (gpt-oss-20b~\cite{openai2025gptoss120bgptoss20bmodel}, qwen-3-32b~\cite{yang2025qwen3technicalreport}) LLMs. Using perturbation-based analysis, we measure how LLMs' attention is distributed across bug report sections and investigate how these patterns differ between successful and unsuccessful repairs. Additionally, we examine whether LLMs attend to the same sections and fine-grained phrases that developers consider most important for repair. Specifically, we investigate the following questions: 

\noindent
\textbf{- RQ1:}
\textit{How is model attention allocated across bug report sections in successful and unsuccessful repairs?} We examine section-level attention using perturbation-based analysis and find that successful repairs consistently prioritize \textit{Bug description} while unsuccessful repairs over-attend to \textit{Version information}.

\noindent
\textbf{- RQ2:}
\textit{How is model attention distributed across specific code and natural-language components within bug report sections?} We identify attention patterns that differentiate repair outcomes using perturbations on natural language and code components within reports, showing that successful repairs rely on \textit{diffused attention} across important information such as stacktrace and test data, while failures often arise from overly \textit{localized attention} on contextual metadata such as library versions.

\noindent
\textbf{- RQ3:}
\textit{How well do model attention patterns align with what developers consider important for bug-fixing?}
Human developers rely on experience-driven intuition to identify the most relevant parts of a bug report during debugging~\cite{4016573, 10.1145/3106237.3106285, 10.1145/3338906.3338947, 10.1109/TSE.2010.63}, but existing APR benchmarks do not capture this information. To fill this gap, we create the first developer attention dataset on a subset of 100 bug reports. Comparing developer annotations with model attention, we show that stronger developer-model attention alignment is significantly associated with successful repairs.

Our findings provide new insights into LLM-based APR systems by revealing where models focus when processing bug reports and how specific attention patterns relate to repair success and failure. These insights can help practitioners design more effective LLM-APR workflows and guide researchers in developing systems that better prioritize diagnostically important information.

The main contributions of this paper are as follows:
(1) We present the first empirical study of attention patterns in LLM-based APR, analyzing how models attend to bug reports during repair.
(2) We characterize both section-level and fine-grained attention patterns across 319 Python and Java bugs, showing how attention allocation differs between successful and unsuccessful repairs.
(3) We evaluate human-model attention alignment by comparing model attention with the information developers consider most important for bug repair.
(4) We present the first annotated developer attention dataset of 100 bug reports for the task of APR.
\section{Background and Motivation}
\begin{figure}[h]
    \centering
    \includegraphics[width=0.9\linewidth]{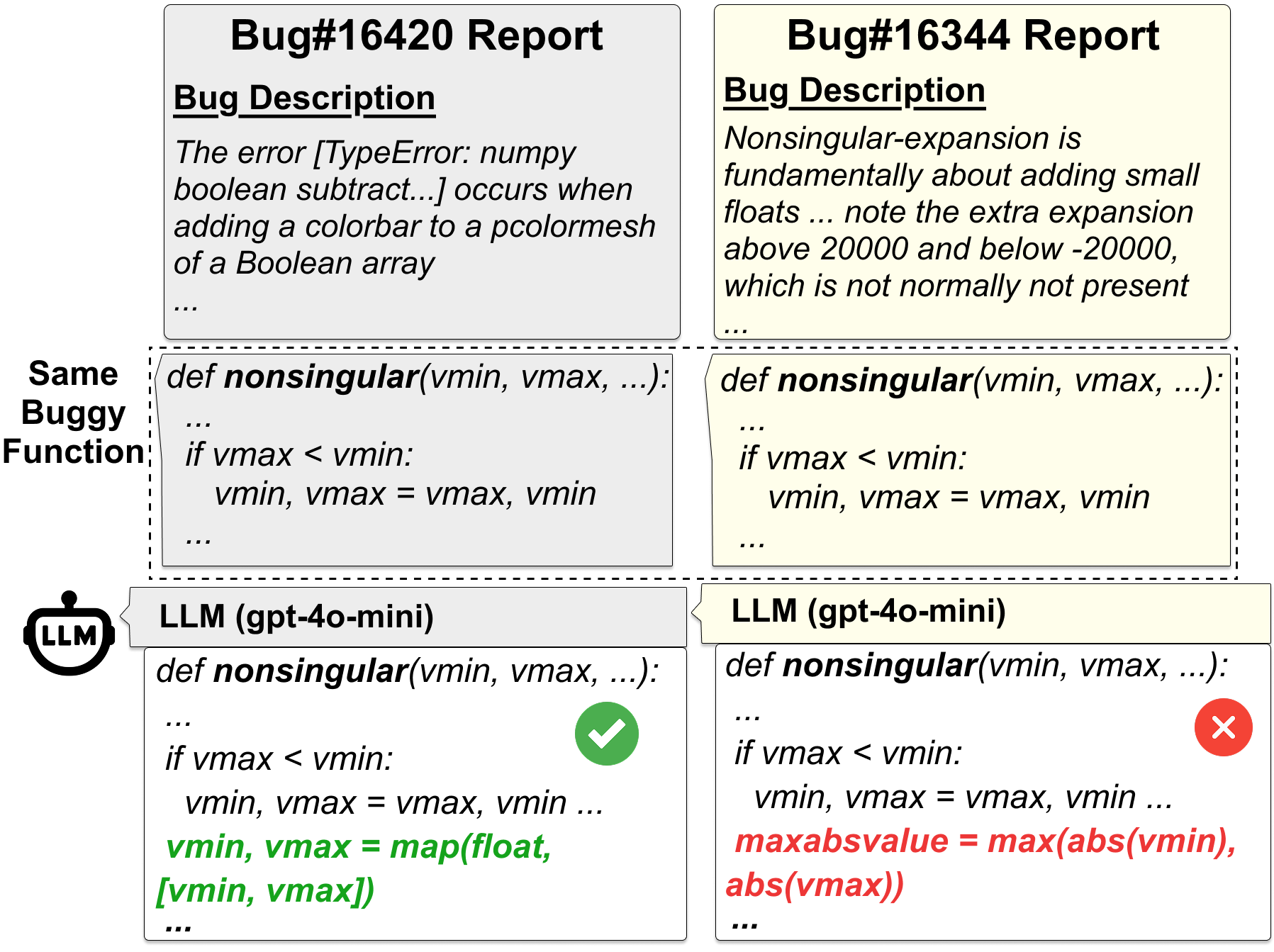}
    \vspace{-0.3cm}
    \caption{Generated patches for bug instances \textit{matplotlib\#16344} and \textit{\#16420}.}
    \vspace{-0.3cm}
    \label{fig:attention_motive}
\end{figure}

Consider the example in Figure~\ref{fig:attention_motive}, which shows two closely related but distinct GUI bugs \#16344 and \#16420 from the \textit{matplotlib} project on GitHub. These two bugs occur within a few months of each other, affect the same function (\texttt{`nonsingular'}) in the same file related to the \texttt{`colorbar'} component, and require the same fix~(\href{https://github.com/matplotlib/matplotlib/commit/05a5db0fec2eced55076736f0b9520641b279ad6}{\textcolor{black}{\underline{link1}}} and \href{https://github.com/matplotlib/matplotlib/commit/5d99e151be80bcb0b3b6d081fd3038330f573d94}{{\textcolor{black}{\underline{link2}}}}).
However, their bug reports point to the same underlying issue in slightly different ways.
Following prior work~\cite{ehsani2025hierarchicalknowledgeinjectionimproving}, we prompt GPT-4o-mini to repair each bug using the same prompt template and model settings, with their corresponding bug reports.
The model successfully generates a correct patch for bug \#16420 but fails to repair bug \#16344.
\textbf{\textit{What drives divergent repair outcomes when the same model is given comparable information for two closely related bugs?}} We cannot answer because we do not know which parts of the bug reports the model attends to and what information it prioritizes during repair, or how this differs between the case where it succeeds and where it fails.

A closer look at the bug reports suggests a potential explanation. Bug \#16420 provides a clear description of the failure (e.g., \texttt{`TypeError'}) along with a reproducible example and an explicit workaround. Bug \#16344 describes the issue in a more indirect way, explaining value ranges without clearly isolating the failure. While both reports contain the necessary information for developers to solve the bugs, they differ in how that information is presented and, therefore, in which parts the model may focus on during repair.

This behavior is not unique to this example.
Such inconsistencies are frequently observed in LLM-based program repair. Even when models are provided with similar information, their repair performance can vary substantially across seemingly similar types of bugs~\cite{ehsani2025detectingpromptknowledgegaps, ehsani2025hierarchicalknowledgeinjectionimproving,ehsani2025makeschatgpteffectivesoftware, 10.1109/ICSE55347.2025.00162}. This unpredictability is widely recognized as one of the central challenges when applying LLMs in practice~\cite{biderman2024lessonstrenchesreproducibleevaluation, laskar-etal-2024-systematic, hochlehnert2025soberlookprogresslanguage, valentin2025incoherenceoraclelessmeasureerror}.

One way to address this challenge is to understand how models attend to different information in the input through \textbf{attention analysis}~\cite{vaswani2023attentionneed}. 
Prior work has examined internal model behavior, such as attention heads across layers~\cite{vaswani2023attentionneed, jain-wallace-2019-attention, serrano-smith-2019-attention, clark-etal-2019-bert, metzger2022attentionflowsgeneraltransformers, mrini-etal-2020-rethinking}, as well as gradient-based signals that estimate which input components most influence predictions~\cite{bastings2020elephantinterpretabilityroomuse, jain-wallace-2019-attention}.
However, directly studying attention within LLM architectures is often infeasible, as many proprietary models do not expose their internal mechanisms. 
Even with open-source access, analyzing attention in large models can still be computationally expensive.

\textbf{Perturbation-based attention analysis} offers a practical alternative. 
\textcolor{black}{Rather than inspecting internal model states or transformer self-attention directly, these methods modify parts of the input and measure how the output changes under deterministic decoding. 
If removing a portion of the input significantly alters the generated output, that component likely played an important role in the model’s decision~\cite{lopardo2024attentionmeetsposthocinterpretability, kou2024large, tang_gint_2023, 9467291}. This motivates our study design: we systematically remove information from bug reports and observe the resulting changes in generated repairs to identify which elements LLMs prioritize during program repair.} Perturbation-based approaches are model-agnostic and can therefore be applied to both proprietary and open-source LLMs. Prior work in software engineering has also shown that perturbation-based methods align well with human reasoning when compared to other methods such as self-attention or gradient-based analysis ~\cite{lopardo2024attentionmeetsposthocinterpretability, kou2024large}.

\section{Methodology}
\noindent
\textbf{\underline{Dataset.}}
We analyze a total of 319 bugs from two widely used APR benchmarks created from real-world GitHub projects: \textit{\swebenchverified}~\cite{jimenez2024swebench} for Python, and \textit{\multiswebench}~\cite{zan2025multiswebenchmultilingualbenchmarkissue} for Java. We focus on Python and Java because they are among the most commonly used languages in software projects~\cite{stackoverflow}. Including both allows us to evaluate whether attention patterns are consistent across different languages.
Because our goal is to analyze attention during patch generation, we adopt \textit{function-level perfect fault localization}~\cite{10.1109/ICSE55347.2025.00162, 8730164, ehsani2025hierarchicalknowledgeinjectionimproving}. Under this setting, the buggy function is given to the model, but it must still identify and repair the faulty lines within it. We manually inspect both benchmarks and retain only single-function bugs, resulting in 248 Python and 71 Java bugs.

\textcolor{black}{Both benchmarks are manually curated by their original authors to include only the bugs that are sufficiently described and solvable by LLMs~\cite{zan2025multiswebenchmultilingualbenchmarkissue, jimenez2024swebench}. This is essential for our study because we aim to understand how attention behaves when relevant information is present, and why some repairs succeed while others fail. In addition, both datasets provide a manually annotated difficulty label for each bug (\textit{Easy}, \textit{Medium}, or \textit{Hard}). Across our dataset, 144 bugs are labeled \textit{Easy}, 164 \textit{Medium}, and 11 \textit{Hard}. The distribution is therefore dominated by easy and medium bugs, with relatively few hard instances. We later use these difficulty annotations to examine whether the observed attention patterns remain consistent after accounting for bug difficulty.}

After manually examining all 319 bug reports, we observed that they consistently contain several common structured sections. We group these sections into the following categories~\cite{li_first_2023, soltani_significance_2020}:
1) \textbf{Bug description}, a description of the symptoms of the bug; 2) \textbf{Reproduction}, steps or instructions to trigger the failure; 3) \textbf{Expected behavior}, a description of the intended or correct system behavior; 4) \textbf{Actual behavior}, a description of the current system behavior; 5) \textbf{Version information}, has information on the specific version of system/tools used when the bug occurs; and 6) \textbf{Additional information}, any supporting details not captured that are relevant to the bug.
Not all bug reports contain every section. However, when present, they consistently fall into these categories.
Across our dataset, \textit{Bug description} appears in all bug reports, \textit{Reproduction} in 133, \textit{Version information} in 98, \textit{Expected behavior} in 97, \textit{Additional information} in 62, and \textit{Actual behavior} in 41. These sections form the basis for our section-level and phrase-level attention analyses.

\noindent
\textbf{{\underline{Models.}}}
We study a combination of proprietary and open-source LLMs.
For proprietary models, we analyze \textit{claude-4-sonnet} because of its state-of-the-art performance on coding benchmarks~\cite{anthropic}.
For open-source models, we use \textit{gpt-oss-20b} and \textit{qwen3-32b}, as they are repeatedly cited among the most capable open-source models for coding~\cite{yang2025qwen3technicalreport, openai2025gptoss120bgptoss20bmodel}.
Studying both sets of models allows us to analyze attention patterns in both commercial and open-source models. We also include both small and large models to see whether attention patterns are consistent across scales.

In \textit{RQ1}, we perform section-level attention analysis across all models to obtain a comparative view of how different LLMs distribute attention over bug report sections. 
In contrast, the perturbation analyses for \textit{RQ2} and \textit{RQ3} are substantially more expensive, as they require masking each component in a bug report and regenerating a patch for every perturbation. Across 319 bugs, this would result in 2,873 perturbations and a large number of model executions. We therefore focus our \textit{RQ2} and \textit{RQ3} analysis for  \textit{qwen3-32b} only. 
We select \textit{qwen3-32b} because it is a strong open-source LLM suitable for multi-lingual coding and is often used as the backbone of several APR systems~\cite{yang2025qwen3technicalreport, yang2025inputreductionenhancedllmbased, lecong2025memoryefficientlargelanguagemodels, akbarpour2025collaborativeagentsautomatedprogram, hu2025tsaprtreesearchframework}.
\textcolor{black}{To assess the generality of our findings, we additionally repeat the RQ2 analysis on the Java subset (71 bugs) of our dataset using \textit{gpt-oss-20b} to see if it exhibits the same overall attention trends observed with \textit{qwen3-32b}.}

For all models, patches are generated using deterministic decoding. We set the temperature to zero and disable reasoning modes when available. This reduces sampling variability and ensures that differences in outputs are primarily due to prompt perturbations.

\noindent
\textbf{\underline{Prompts.}}
\textcolor{black}{All prompts use a standardized template designed for program repair~\cite{10.1109/ICSE55347.2025.00162, ehsani2025hierarchicalknowledgeinjectionimproving}. Each prompt begins with high-level instructions describing the repair task, followed by the buggy function to be fixed, and finally the full bug report. We use this standardized prompt template across all models to reduce confounding factors introduced by prompt structure variations. The detailed prompt template is provided in our replication package~\cite{rep_package}.}

\subsection{Research Questions}

\subsubsection{RQ1: How is model attention allocated across bug report sections in successful and unsuccessful repairs?}
\label{rq1}
\textcolor{black}{We first compare repair performance across models and bug difficulty levels. To better understand successful and unsuccessful repairs, we compare all patches generated by each model against the corresponding ground-truth developer patches using the CodeBLEU similarity metric~\cite{ren2020codebleumethodautomaticevaluation}.}

Then, to examine whether repair success depends on which sections the model prioritizes when generating a patch, we use perturbation-based attention analysis. Specifically, we consider each bug report as a set of six distinct sections: \textit{Bug description}, \textit{Reproduction}, \textit{Expected behavior}, \textit{Actual behavior}, \textit{Version Information}, and \textit{Additional information}. Each section is treated as a feature for perturbation that can be either \emph{kept} or \emph{masked}. We identify these sections using the headers present in the markdown of each report. 

Following prior work~\cite{kou2024large}, we define a binary mask $\mathbf{z}\in\{0,1\}^m$ over the $m$ sections of a bug report, where $z_i{=}1$ indicates that section~$i$ is removed. For each mask $\mathbf{z}$, we regenerate a patch and measure how much it differs from the patch generated using the full report, i.e.,  $\mathbf{z}=\mathbf{0}$.
To estimate the contribution of each section, we apply Kernel SHAP over coalitions of masked sections~\cite{10.1145/3763230, 10.5555/3295222.3295230}, using CodeBLEU similarity~\cite{ren2020codebleumethodautomaticevaluation} to quantify changes in the generated patch~\cite{kou2024large}.
A larger change indicates that the removed section had a greater influence on the model’s output~\cite{kou2024large}.

SHAP (SHapley Additive exPlanations)~\cite{10.5555/3295222.3295230} provides a Shapley value for each section that reflects its marginal contribution to the observed output difference. We convert absolute SHAP magnitudes into an attention score by normalizing them within each bug report so that the scores sum to 100 across sections. 
In practice, this corresponds to an occlusion-style attribution method: if removing section $i$ substantially changes the produced patch, then that section receives higher attention~\cite{kou2024large}. We then aggregate SHAP-derived attention scores across bugs and compare distributions between successful and unsuccessful repairs. To quantify the magnitude and significance of the differences, we use Cliff’s $\delta$ effect size measurement and the Mann-Whitney U test ($\alpha=0.05$), following best practices of statistical reporting~\cite{kitchenham_robust_2017}.

\subsubsection{RQ2: How is model attention distributed across specific code and natural-language components within bug report sections?}
\label{rq2}
To analyze fine-grained attention patterns within sections, we break down their content into natural language and code components. \textcolor{black}{These components are automatically extracted using regex-based heuristics and manually verified by the first author.}

To identify \textbf{natural language} components in each section, we perform sentence-level segmentation using the NLTK sentence tokenizer~\cite{nltk}. \textcolor{black}{Each sentence is associated with the Markdown section header under which it appears,} yielding the following categories:
\begin{itemize}[leftmargin=*]
\item \textbf{NL: Description}, sentences describing the overall bug or failure;
\item \textbf{NL: Version information}, sentences describing system versions or environment configurations;
\item \textbf{NL: Expected behavior}, sentences describing the intended program behavior;
\item \textbf{NL: Actual behavior}, sentences describing the observed incorrect behavior;
\item \textbf{NL: Reproduction}, sentences describing how to reproduce the bug; and
\item \textbf{NL: Additional information}, sentences describing any supplementary details related to the bug.
\end{itemize}

\textcolor{black}{To identify \textbf{code-related} information, we apply regex (e.g., triple backticks, log formats) tailored to the two languages in our dataset, Python and Java.} Using this approach, we derive four categories:
\begin{itemize}[leftmargin=*]
\item \textbf{Code: Stacktrace} corresponding to the runtime error traces that often indicate the failure location;
\item \textbf{Code: Test} snippets representing the reproduction tests or assertions used to demonstrate the bug;
\item \textbf{Code: Class and method} definitions that illustrate relevant portions of the system’s implementation; and 
\item \textbf{Code: Import and variable} declarations that reference external APIs or system interfaces.
\end{itemize}
\textcolor{black}{We manually verify all identified components to ensure accurate classification.} Each component is then treated as an independent information unit for evaluating its influence on repair behavior.

\begin{figure}[h]
    \centering
    \includegraphics[width=0.9\linewidth]{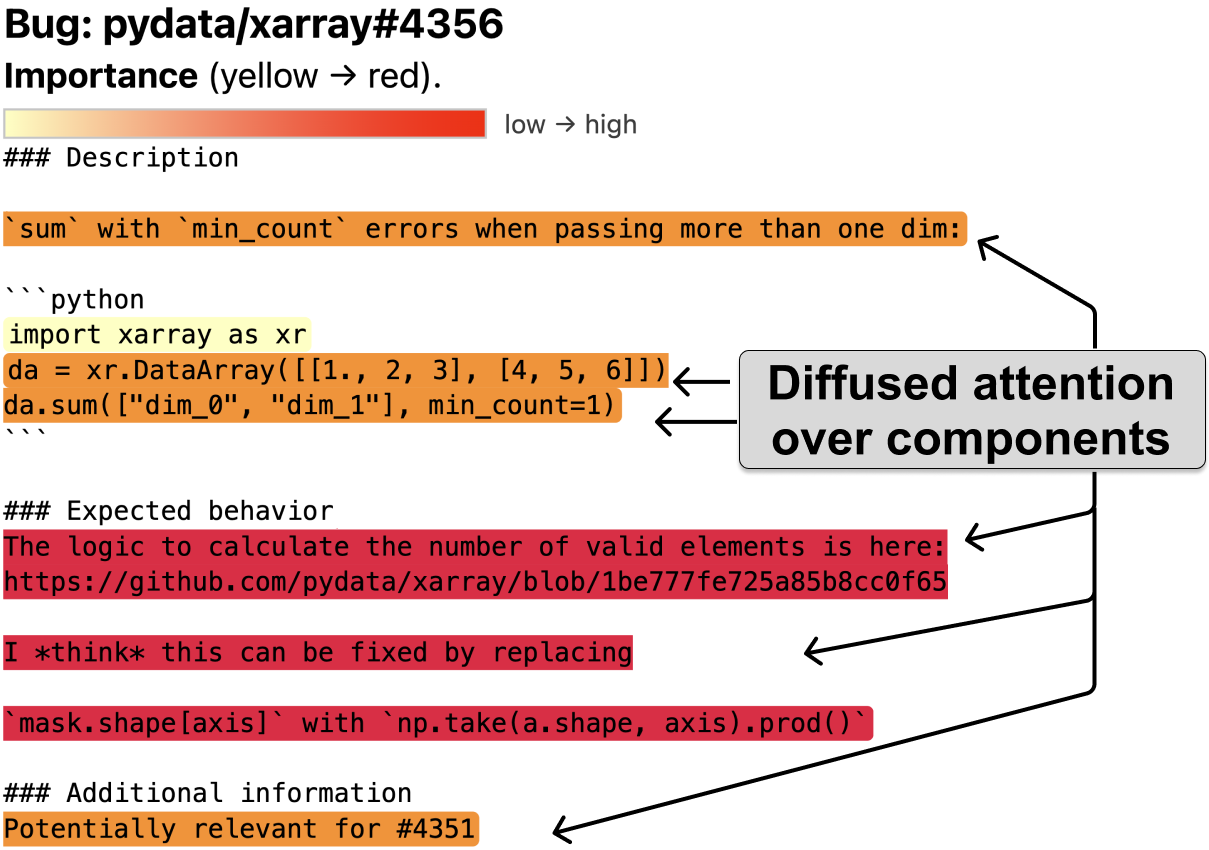}
    \vspace{-0.3cm}
    \caption{Example of Diffused Attention Pattern in LLMs.}
    \label{fig:diffused_attention}
    \vspace{-0.3cm}
\end{figure}

For each bug instance, we denote the 
output obtained by retaining all sections of a bug report as $O_{base}$. We then perform perturbation analysis within a section by masking one component at a time, keeping the remaining components unchanged. This perturbed bug report is then passed to the model to generate a new patch.
To quantify the influence of each component, we measure the semantic difference between the original output ($O_{base}$) and the perturbed output ($O_{pert}$) using UniXcoder~\cite{guo2022unixcoderunifiedcrossmodalpretraining}. UniXcoder provides code-aware embeddings that incorporate syntactic information through abstract syntax tree representations.
The importance score of component $c$ is defined as the change in similarity between $O_{base}$ and $O_{pert}$. Intuitively, if masking a component causes a large semantic change in the generated patch, the removed information is considered to have a strong influence on the model’s repair behavior. Aggregating these importance scores across components in different sections of a bug report creates an attention pattern distribution map for each bug report. 

\begin{figure}[h]
    \centering
    \includegraphics[width=0.95\linewidth]{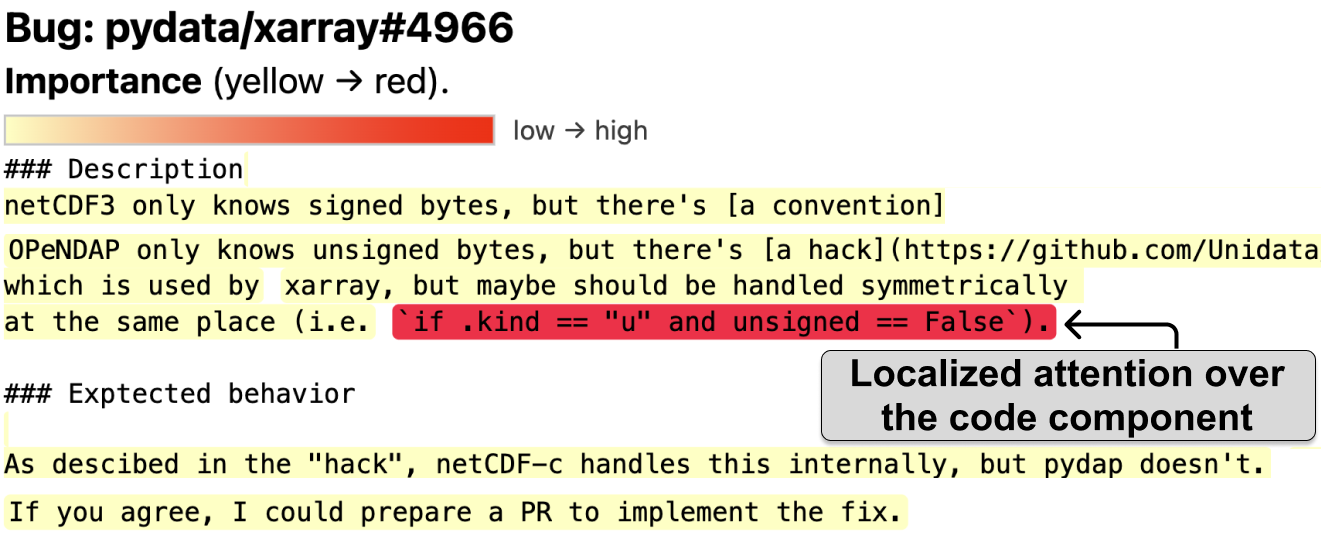}
    \vspace{-0.4cm}
    \caption{Example of Localized Attention Pattern in LLMs.}
    \vspace{-0.35cm}
    
    \label{fig:localized_attention}
\end{figure}

\begin{figure}[h]
    \centering
    \includegraphics[width=0.85\linewidth]{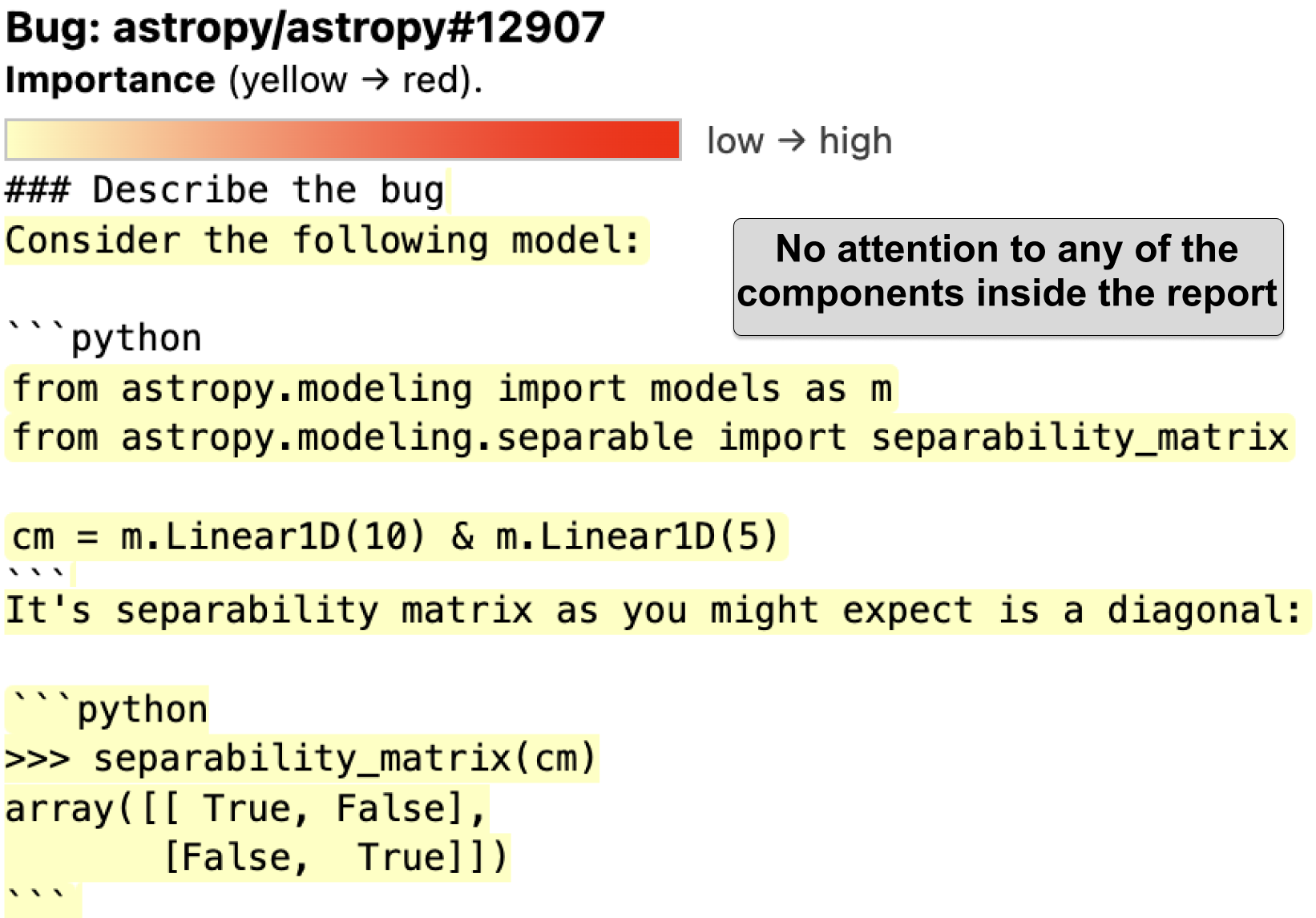}
    \vspace{-0.4cm}
    
    \caption{Example of No-attention Pattern in LLMs.}
    \vspace{-0.35cm}
    
    \label{fig:no_attention}
\end{figure}

We visualize component-level importance scores by mapping them back onto the original bug report. Each component is highlighted using a color scale from yellow to red, with warmer colors indicating greater model attention (see Figure~\ref{fig:diffused_attention}). We then analyze these visualizations to identify recurring attention patterns across bugs. Through this process, we identify three attention structures:

\begin{itemize}[leftmargin=*]
\item \textbf{Diffused}, where model attention is distributed across multiple components (e.g., Figure~\ref{fig:diffused_attention}, where attention is distributed across \textit{NL: Expected Behavior}, \textit{Code: Class and method}, \textit{NL: Description}, and \textit{NL: Additional information});
\item \textbf{Localized}, where the model focuses heavily on a single component while ignoring other information (e.g., Figure~\ref{fig:localized_attention}, where model attention is only on \textit{Code: Class and method}); 
\item \textbf{No-attention}, where masking any of the components produces no changes in the generated output (e.g., Figure~\ref{fig:no_attention}).
\end{itemize}

We also investigate which components (e.g., stacktraces or natural language sections) receive the strongest attention in successful bug repair scenarios. To evaluate the relationship between attention patterns and repair success, we construct contingency tables relating the presence of each attention behavior to patch correctness. We apply Fisher’s Exact Test due to the binary nature of the patterns~\cite{upton_fishers_1992}. Fisher’s test is appropriate for our data because it computes exact probabilities and performs reliably with relatively small sample sizes~\cite{kim_statistical_2017}.
To control for multiple hypothesis testing, we apply the Benjamini-Hochberg False Discovery Rate (FDR) correction~\cite{benjamini_controlling_1995}, following recommended statistical reporting in empirical software engineering~\cite{kitchenham_robust_2017}.
In addition, we use logistic regression to report odds ratios and confidence intervals to quantify the magnitude and direction of each attention pattern’s association with repair success.
\textcolor{black}{To account for bug difficulty as a potential confounding factor, we also repeat the logistic regression analysis by including the benchmark difficulty labels (\textit{Easy}, \textit{Medium}, \textit{Hard}) as control variables, with \textit{Easy} used as the reference category. This allows us to estimate the association between each attention pattern and repair success after adjusting for differences in bug difficulty.} 

\subsubsection{RQ3: How well do model attention patterns align with what developers consider important for bug-fixing?}
While prior work has shown that the information developers focus on strongly affects their ability to understand and fix bugs~\cite{10.1145/3106237.3106285, 10.1145/3338906.3338947, 10.1109/TSE.2010.63}, existing APR benchmarks do not provide information about which parts of a bug report developers consider most useful during repair. 
To fill this gap, we construct the first developer attention dataset. From our dataset of 319 bugs, we draw a stratified sample of 100 bug reports (50 Python and 50 Java). 
This sample size balances sufficient statistical power with approximately 95\% confidence and a ±8\% margin of error~\cite{cochran1977sampling}, with the substantial manual effort required for detailed attention annotation.

\textcolor{black}{We recruited four experienced software developers (with 5+ years of programming experience in Python and Java) to manually annotate these 100 bug reports. Each bug report is annotated by one developer, each labeling 25 bugs. 
Using insights from prior studies on how developers identify relevant information from software artifacts~\cite{ChatterjeeJSS, 10.1145/3510003.3510108, 10.1109/TSE.2010.63, soltani_significance_2020}, we designed the annotation study to capture which bug report content developers consider most important for repair.
For each bug report, \textbf{(1)} the developers rate the importance of each section on a five-point Likert scale (1\,=\,not important, 5\,=\,very important)~\cite{likert1932technique}, \textbf{(2)} select the top two most important sections, and \textbf{(3)} provide exact-quote key phrases within the selected sections that capture essential information.}

\textcolor{black}{Each bug report is annotated by a single experienced developer because the goal of the study is to capture subjective notions of importance during bug repair to test alignment of LLMs and developers rather than establish a single objective ground truth to test alignment between human developers. Different developers may prioritize different information when debugging. However, to reduce excessive subjectivity, before annotation began, we prepared detailed annotation guidelines and conducted a pilot session to refine the procedure and ensure a consistent understanding of the task. In addition, the first author annotated 40 sampled bugs (10 from each annotator) and measured agreement overlap with the original annotations. We see an average Spearman correlation~\cite{james2026countingconsensusselectingright} of 0.817 for section-level ratings, a Cohen’s Kappa~\cite{keppaArticle} of 0.77 for Top-2 selections, and an average Top-2 overlap of 1.475/2 sections, which all suggest strong consistency across the overlapped annotations.
Annotations are collected using a custom Streamlit-based~\cite{streamlit} interface developed for this study. More information on annotation instructions and the tool is included in our replication package~\cite{rep_package}.}

We compare these annotations with model attention at two levels of granularity: a) section-level, and b) phrase-level.

\textbf{a) Section-level}: We derive section-level model attention by aggregating the importance scores from \textit{RQ2} within each bug-report section. This yields a normalized attention vector $\{a_s\}$ over sections, where $a_s$ reflects the total importance assigned to information units appearing under section $s$.
We measure alignment using three metrics. First, we compute \textit{Hit@1}, the overlap between the developer-selected top section $H_{\text{top1}}$ and the model’s top attended section $M_{\text{top1}}$ ($|H_{\text{top1}} \cap M_{\text{top1}}|$), and \textit{Hit@2}, the overlap between the developer-selected top-two sections $H_{\text{top2}}$ and the model’s top-two attended sections $M_{\text{top2}}$, i.e., $|H_{\text{top2}} \cap M_{\text{top2}}|$.
Then, we compute the \textit{Spearman rank correlation} between the developer section importance ratings $\{r_s\}$ and the model section attention scores $\{a_s\}$ for each bug report.
\textbf{b) Phrase-level}: To compare developer-identified key phrases to model attention, we use the results in \textit{RQ2} to obtain a phrase-level attention score over bug reports. We map each developer-provided quote to its corresponding token span, yielding a set of key phrases $K$. We then measure phrase-level alignment using San Martino’s token overlap metrics~\cite{kou2024large}, which quantify the overlap between the set of top-$k$ phrases attended by the model and the developer-annotated key phrases. Specifically, we report Precision@k, Recall@k, and F1@k for k $=10, 20$.

\section{Results and Discussion}

\subsection{RQ1: How is model attention allocated across bug report sections in successful and unsuccessful repairs?}
\textbf{\underline{Results. }}\textcolor{black}{As shown in Table~\ref{tab:solved_distribution}, \textit{claude-4} achieves the highest fix rate (65\%), followed by \textit{gpt-oss} (51\%) and \textit{qwen-3} (40\%).
Across all three models, repair rates are consistently higher for \textit{Easy} bugs (48-71\%) than for \textit{Medium} bugs (34-62\%). The \textit{Hard} category exhibits 36\% to 55\% repair rates, but it contains only 11 bugs and does not provide sufficient statistical power for reliable conclusions. A chi-square test of independence with correction~\cite{mchugh_chi-square_2013} finds no statistically significant association between benchmark difficulty and repair success ($p=0.06$), suggesting that benchmark difficulty alone does not fully explain repair outcomes.}

\textcolor{black}{Across our models, successful repairs exhibit high similarity to the ground-truth patches (median CodeBLEU: 91\% for \textit{claude-4}, 85\% for \textit{gpt-oss}, and 84\% for \textit{qwen-3}), indicating that correct repairs closely match the developer implementations despite syntactic differences such as variable names or code formatting.
In contrast, failed repairs show substantially lower similarity to the ground-truth patches (mean CodeBLEU: 30\% for \textit{claude-4}, 29\% for \textit{gpt-oss}, and 31\% for \textit{qwen-3}). This indicates that failures are generally different from the ground-truth implementations rather than near-correct fixes. Our manual inspection showed that failed repairs do not capture the intended bug behavior described in the reports. We discuss representative examples of these failures throughout the discussions of each \textit{RQ}.}

\begin{table}[t]
\caption{\textcolor{black}{Fix-rate of models across our dataset by difficulty.}}
\vspace{-0.3cm}
\centering
\footnotesize
\resizebox{\columnwidth}{!}{%
\begin{tabular}{ll|c|c|c|c}
\toprule
\textbf{Model} & \textbf{Data} & \textbf{Overall} & \textbf{Easy} & \textbf{Medium} & \textbf{Hard} \\
\midrule
\multirow{3}{*}{\texttt{qwen-3-32b}}
 & Python &
\begin{tabular}[c]{@{}c@{}}96/248\\(38.7\%)\end{tabular} &
\begin{tabular}[c]{@{}c@{}}59/124\\(47.6\%)\end{tabular} &
\begin{tabular}[c]{@{}c@{}}37/120\\(30.8\%)\end{tabular} &
\begin{tabular}[c]{@{}c@{}}0/4\\(0.0\%)\end{tabular} \\
\cline{2-6}
 & Java &
\begin{tabular}[c]{@{}c@{}}34/71\\(47.8\%)\end{tabular} &
\begin{tabular}[c]{@{}c@{}}10/20\\(50.0\%)\end{tabular} &
\begin{tabular}[c]{@{}c@{}}18/44\\(40.9\%)\end{tabular} &
\begin{tabular}[c]{@{}c@{}}6/7\\(85.7\%)\end{tabular} \\
\cline{2-6}
 & \textbf{All} &
\begin{tabular}[c]{@{}c@{}}\textbf{130/319}\\\textbf{(40.7\%)}\end{tabular} &
\begin{tabular}[c]{@{}c@{}}\textbf{69/144}\\\textbf{(47.9\%)}\end{tabular} &
\begin{tabular}[c]{@{}c@{}}\textbf{55/164}\\\textbf{(33.5\%)}\end{tabular} &
\begin{tabular}[c]{@{}c@{}}\textbf{6/11}\\\textbf{(54.5\%)}\end{tabular} \\
\midrule
\multirow{3}{*}{\texttt{gpt-oss-20b}}
 & Python &
\begin{tabular}[c]{@{}c@{}}133/248\\(53.6\%)\end{tabular} &
\begin{tabular}[c]{@{}c@{}}72/124\\(58.1\%)\end{tabular} &
\begin{tabular}[c]{@{}c@{}}60/120\\(50.0\%)\end{tabular} &
\begin{tabular}[c]{@{}c@{}}1/4\\(25.0\%)\end{tabular} \\
\cline{2-6}
 & Java &
\begin{tabular}[c]{@{}c@{}}32/71\\(45.1\%)\end{tabular} &
\begin{tabular}[c]{@{}c@{}}10/20\\(50.0\%)\end{tabular} &
\begin{tabular}[c]{@{}c@{}}19/44\\(43.2\%)\end{tabular} &
\begin{tabular}[c]{@{}c@{}}3/7\\(42.9\%)\end{tabular} \\
\cline{2-6}
 & \textbf{All} &
\begin{tabular}[c]{@{}c@{}}\textbf{165/319}\\\textbf{(51.7\%)}\end{tabular} &
\begin{tabular}[c]{@{}c@{}}\textbf{82/144}\\\textbf{(56.9\%)}\end{tabular} &
\begin{tabular}[c]{@{}c@{}}\textbf{79/164}\\\textbf{(48.2\%)}\end{tabular} &
\begin{tabular}[c]{@{}c@{}}\textbf{4/11}\\\textbf{(36.4\%)}\end{tabular} \\
\midrule
\multirow{3}{*}{\texttt{claude-4-sonnet}}
 & Python &
\begin{tabular}[c]{@{}c@{}}165/248\\(66.5\%)\end{tabular} &
\begin{tabular}[c]{@{}c@{}}92/124\\(74.2\%)\end{tabular} &
\begin{tabular}[c]{@{}c@{}}72/120\\(60.0\%)\end{tabular} &
\begin{tabular}[c]{@{}c@{}}1/4\\(25.0\%)\end{tabular} \\
\cline{2-6}
 & Java &
\begin{tabular}[c]{@{}c@{}}43/71\\(60.6\%)\end{tabular} &
\begin{tabular}[c]{@{}c@{}}10/20\\(50.0\%)\end{tabular} &
\begin{tabular}[c]{@{}c@{}}30/44\\(68.2\%)\end{tabular} &
\begin{tabular}[c]{@{}c@{}}3/7\\(42.9\%)\end{tabular} \\
\cline{2-6}
 & \textbf{All} &
\begin{tabular}[c]{@{}c@{}}\textbf{208/319}\\\textbf{(65.2\%)}\end{tabular} &
\begin{tabular}[c]{@{}c@{}}\textbf{102/144}\\\textbf{(70.8\%)}\end{tabular} &
\begin{tabular}[c]{@{}c@{}}\textbf{102/164}\\\textbf{(62.2\%)}\end{tabular} &
\begin{tabular}[c]{@{}c@{}}\textbf{4/11}\\\textbf{(36.4\%)}\end{tabular} \\
\bottomrule
\end{tabular}%
}
\label{tab:solved_distribution}
\vspace{-0.5cm}
\end{table}

Table~\ref{tab:attention_effect_heatmap} reports attention differences between solved and unsolved repairs by section for each model using Cliff’s $\delta$. Positive values indicate that attention scores tend to be higher for successful repairs, while negative values indicate higher attention for unsuccessful repairs. 
Across all models, \textit{Bug description} and \textit{Version information} sections exhibit the largest differences. The \textit{Bug description} section shows large positive effect sizes of \textbf{0.54,} \textbf{0.51}, and \textbf{0.50} for \textit{claude-4}, \textit{gpt-oss}, and \textit{qwen-3}, respectively, indicating that successful repairs tend to allocate more attention to the description of the problem. In contrast, the \textit{Version information} section shows large negative effect sizes of \textbf{-0.53}, \textbf{-0.20}, and \textbf{-0.42} for each model, suggesting that unsuccessful repairs tend to allocate more attention to environmental metadata.
Despite differences in overall repair rates across models, the direction and magnitude of the attention differences are consistent. This suggests that similar attention patterns distinguish successful from unsuccessful repairs across models.
These same patterns also appear across both Python and Java bugs, suggesting stability across these programming languages. 

Other sections, including \textit{Reproduction}, \textit{Actual Behavior}, \textit{Expected Behavior}, and \textit{Additional Information} show small or inconsistent effect sizes across models, indicating limited differences in attention allocation between successful and unsuccessful repairs. 

\begin{table}[h]
\caption{Comparing section attention between solved and unsolved bugs using Effect-size (Cliff's $\delta$). Top-2 biggest differences in each model are highlighted (red: top-1, orange: top-2). Bold* denotes statistical significance ($p < 0.05$) with the Mann-Whitney U test.}
\vspace{-0.3cm}
\centering
\footnotesize
\setlength{\tabcolsep}{1pt}
\begin{tabular}{lccc ccc ccc}
\toprule
& \multicolumn{3}{c}{\texttt{claude-4-sonnet}}
& \multicolumn{3}{c}{\texttt{gpt-oss-20b}}
& \multicolumn{3}{c}{\texttt{qwen-3-32b}} \\

\cmidrule(lr){2-4}
\cmidrule(lr){5-7}
\cmidrule(lr){8-10}

\textbf{Section}
& \textbf{All} & \textbf{Python} & \textbf{Java}
& \textbf{All} & \textbf{Python} & \textbf{Java}
& \textbf{All} & \textbf{Python} & \textbf{Java} \\

\midrule
\begin{tabular}[c]{@{}l@{}}Bug\\ description\end{tabular} & \cellcolor{effectLarge}\textbf{0.54*} & \textbf{0.53*} & \textbf{0.55*} & \cellcolor{effectLarge}\textbf{0.51*} & \textbf{0.51*} & \textbf{0.49*} & \cellcolor{effectLarge}\textbf{0.50*} & \textbf{0.48*} & \textbf{0.52*} \\
\begin{tabular}[c]{@{}l@{}}Version\\ information\end{tabular} & \cellcolor{effectMedium}\textbf{$-$0.53*} & \textbf{$-$0.52*} & \textbf{$-$0.48*} & \cellcolor{effectMedium}$-0.20$ & $-0.22$ & $-0.13$ & \cellcolor{effectMedium}\textbf{$-$0.42*} & \textbf{$-$0.47*} & $-$0.32 \\
Reproduction & $-0.08$ & $-0.12$ & 0.08 & 0.09 & 0.15 & $-0.05$ & 0.14 & 0.05 & 0.35 \\
\begin{tabular}[c]{@{}l@{}}Actual\\ behavior\end{tabular} & $-0.30$ & $-0.29$ & $-$1.00 & $-0.03$ & $-0.20$ & 0.75 & $-0.08$ & $-0.02$ & 0.0 \\
\begin{tabular}[c]{@{}l@{}}Expected\\ behavior\end{tabular} & $-0.21$ & $-0.25$ & 0.06 & 0.00 & 0.05 & $-0.04$ & $-0.04$ & $-0.13$ & 0.22 \\
\begin{tabular}[c]{@{}l@{}}Additional\\ information\end{tabular} & $0.16$ & 0.12 & 0.34 & $-0.04$ & 0.04 & $-0.25$ & 0.18 & 0.15 & 0.20 \\ \hline
\end{tabular}%
\label{tab:attention_effect_heatmap}
\vspace{-0.3cm}
\end{table}

\begin{table*}[!t]
\centering
\footnotesize
\setlength{\tabcolsep}{1pt}
\caption{Associations between attention patterns and repair success using Fisher’s Exact Test with Benjamini-Hochberg FDR correction (FDR $q$). Odds Ratios (OR) are reported with bootstrap Confidence Intervals (CI). The column \textit{Influence} indicates the direction of the association based on the OR value (OR $>$ 1 more in successful; OR $<$ 1 more in unsuccessful repairs). Rows with statistically significant associations ($q < 0.05$) are marked with \textbf{*} and shaded according to the direction of their effect: green indicates a positive association with successful repairs, while red indicates a negative association with repair success.}
\vspace{-0.4cm}
\label{tab:attn_or_fisher}

\begin{tabularx}{\linewidth}{@{}L D R r r c R r r c R r r c@{}}
\toprule

& & \multicolumn{4}{c}{\textbf{Overall}}
& \multicolumn{4}{c}{\textbf{Python}}
& \multicolumn{4}{c}{\textbf{Java}} \\

\cmidrule(lr){3-6}
\cmidrule(lr){7-10}
\cmidrule(lr){11-14}

\textbf{Pattern} & \textbf{Influence}
& \textbf{OR [CI]} & \textbf{$p$-value} & \textbf{FDR $q\ \ $} & \textbf{Count}
& \textbf{OR [CI]} & \textbf{$p$-value} & \textbf{FDR $q\ \ $} & \textbf{Count}
& \textbf{OR [CI]} & \textbf{$p$-value} & \textbf{FDR $q\ \ $} & \textbf{Count} \\

\midrule

\multicolumn{14}{@{}l@{}}{\textbf{\underline{Attention Structure}}} \\
\addlinespace[2pt]

No-attention & \neutraldir
& 1.0 [1.0,1.0] & $8.87{\times}10^{-1}$ & $9.60{\times}10^{-1}$ & 64
& 1.0 [1.0,1.0] & $1.97{\times}10^{-1}$ & $2.56{\times}10^{-1}$ & 51
& 1.0 [1.0,1.0] & $6.65{\times}10^{-2}$ & $1.44{\times}10^{-1}$ & 13 \\

\rowcolor{pos}
\textbf{Diffused*} & \up
& \textbf{2.07 [1.68,2.52]} & $3.71{\times}10^{-15}$ & $2.41{\times}10^{-14}$ & 117
& 1.66 [1.28,2.13] & $4.63{\times}10^{-9}$ & $3.01{\times}10^{-8}$ & 84
& 1.86 [1.53,2.21] & $7.99{\times}10^{-8}$ & $1.00{\times}10^{-6}$ & 33 \\

\rowcolor{neg}
\textbf{Localized*} & \down
& \textbf{0.40 [0.33,0.50]} & $7.38{\times}10^{-16}$ & $9.60{\times}10^{-15}$ & 138
& 0.40 [0.32,0.53] & $9.86{\times}10^{-12}$ & $1.28{\times}10^{-10}$ & 113
& 0.65 [0.54,0.81] & $7.25{\times}10^{-5}$ & $4.71{\times}10^{-4}$ & 25 \\

\midrule

\multicolumn{14}{@{}l@{}}{\textbf{\underline{Focus Targets}}} \\
\addlinespace[2pt]

\rowcolor{pos}
\textbf{NL:Description*} & \up
& 1.42 [1.09,1.85] & $1.05{\times}10^{-2}$ & $2.79{\times}10^{-2}$ & 129
& 1.12 [0.85,1.48] & $4.22{\times}10^{-1}$ & $4.57{\times}10^{-1}$ & 95
& 1.57 [1.24,1.93] & $3.37{\times}10^{-4}$ & $1.46{\times}10^{-3}$ & 34 \\

\textbf{NL:}Expected behavior & \up
& 1.31 [1.06,1.65] & $6.37{\times}10^{-2}$ & $1.04{\times}10^{-1}$ & 21
& 1.30 [1.05,1.63] & $4.83{\times}10^{-2}$ & $1.57{\times}10^{-1}$ & 18
& 1.04 [0.94,1.15] & $6.04{\times}10^{-1}$ & $7.14{\times}10^{-1}$ & 3 \\

\textbf{NL:}Reproduction & \up
& 1.23 [0.96,1.59] & $3.04{\times}10^{-2}$ & $5.65{\times}10^{-2}$ & 24
& 1.15 [0.88,1.49] & $1.51{\times}10^{-1}$ & $2.56{\times}10^{-1}$ & 20
& 1.13 [1.03,1.26] & $4.77{\times}10^{-2}$ & $1.24{\times}10^{-1}$ & 4 \\

\textbf{NL:}Actual behavior & \up
& 1.06 [0.90,1.24] & $1.26{\times}10^{-1}$ & $1.63{\times}10^{-1}$ & 7
& 1.10 [0.93,1.28] & $1.12{\times}10^{-1}$ & $2.56{\times}10^{-1}$ & 7
& 1.00 [1.00,1.00] & $1.00$ & $1.00$ & 0 \\

\textbf{NL:}Additional information & \down
& 0.98 [0.78,1.24] & $1.00$ & $1.00$ & 12
& 0.94 [0.78,1.11] & $1.00$ & $1.00$ & 9
& 1.05 [0.89,1.24] & $6.04{\times}10^{-1}$ & $7.14{\times}10^{-1}$ & 3 \\

\rowcolor{neg}
\textbf{NL:Version information*} & \down
& 0.61 [0.49,0.77] & $1.41{\times}10^{-2}$ & $3.05{\times}10^{-2}$ & 27
& 0.72 [0.58,0.89] & $1.66{\times}10^{-1}$ & $2.56{\times}10^{-1}$ & 21
& 0.81 [0.69,0.93] & $2.56{\times}10^{-2}$ & $8.33{\times}10^{-2}$ & 6 \\

\cmidrule(r){1-1}

\rowcolor{pos}
\textbf{Code:Stacktrace*} & \up
& 1.41 [1.11,1.81] & $3.61{\times}10^{-3}$ & $1.57{\times}10^{-2}$ & 54
& 1.36 [1.06,1.74] & $5.17{\times}10^{-3}$ & $2.24{\times}10^{-2}$ & 41
& 1.12 [0.94,1.33] & $3.62{\times}10^{-1}$ & $5.22{\times}10^{-1}$ & 13 \\

\textbf{Code:}Classes and methods & \up
& 1.26 [0.94,1.69] & $7.26{\times}10^{-2}$ & $1.05{\times}10^{-1}$ & 45
& 1.19 [0.88,1.61] & $2.60{\times}10^{-1}$ & $3.08{\times}10^{-1}$ & 35
& 1.14 [0.94,1.35] & $1.78{\times}10^{-1}$ & $2.89{\times}10^{-1}$ & 10 \\

\rowcolor{pos}
\textbf{Code:Test*} & \up
& 1.16 [1.03,1.33] & $1.07{\times}10^{-2}$ & $2.79{\times}10^{-2}$ & 5
& 1.07 [1.00,1.19] & $1.48{\times}10^{-1}$ & $2.56{\times}10^{-1}$ & 2
& 1.11 [1.00,1.24] & $1.05{\times}10^{-1}$ & $1.94{\times}10^{-1}$ & 3 \\

\textbf{Code:}Imports and variables & \up
& 1.16 [0.89,1.53] & $1.81{\times}10^{-1}$ & $2.14{\times}10^{-1}$ & 43
& 1.15 [0.88,1.50] & $1.84{\times}10^{-1}$ & $2.56{\times}10^{-1}$ & 34
& 1.06 [0.91,1.25] & $7.29{\times}10^{-1}$ & $7.89{\times}10^{-1}$ & 9 \\

\bottomrule
\end{tabularx}
\vspace{-0.3cm}
\end{table*}

\noindent
\textbf{\underline{Discussion. }}Our findings suggest that successful repairs rely more heavily on the \textit{Bug description} section, which provides the most direct explanation of the failure. Bug descriptions often summarize the underlying issue in natural language and highlight key symptoms of the bug, giving the models clearer signals about what behavior needs to be corrected.

\begin{figure}[h]
    \centering
    \includegraphics[width=0.85\linewidth]{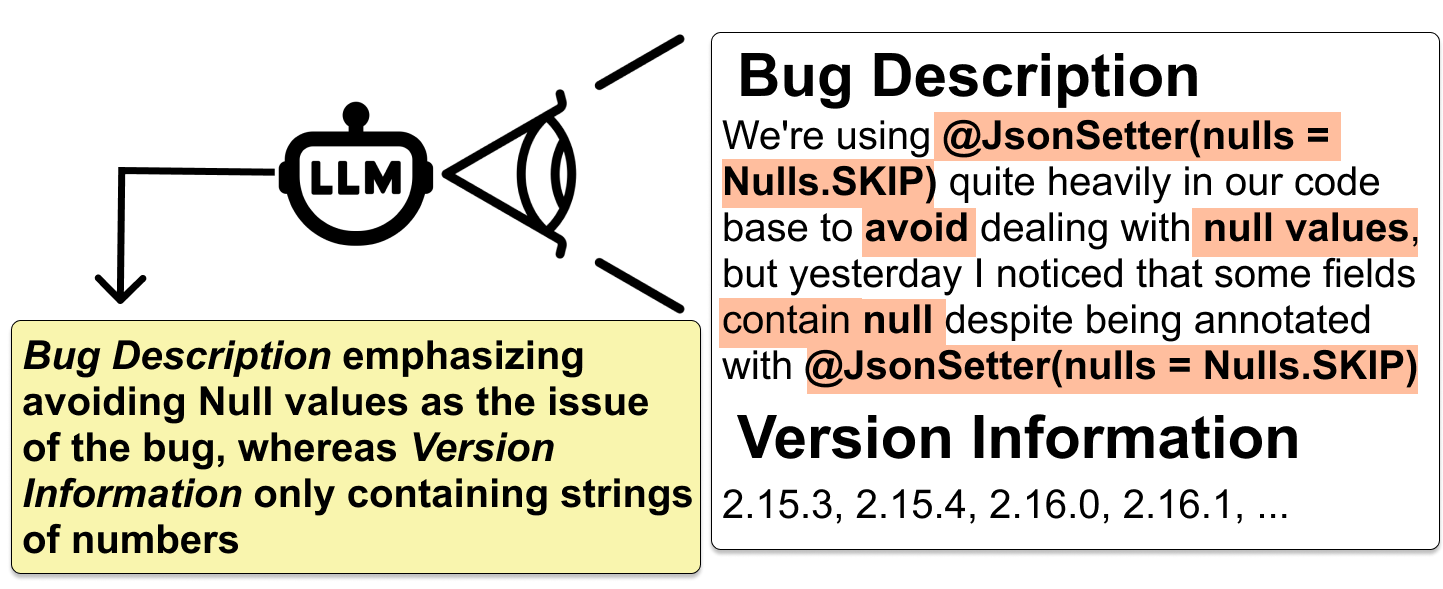}
    \vspace{-0.3cm}
    \caption{Example of Bug Description and Version Information Sections in Project \textit{jackson-databind\#4469}.}
    \label{fig:desc_version}
    \vspace{-0.4cm}
\end{figure}

In contrast, the \textit{Version information} section often contains environmental metadata such as version numbers or system configurations. While this information can help developers reproduce issues in different versions of the code, it can rarely provide actionable insights into the cause of the bug to LLMs.
\textcolor{black}{Figure~\ref{fig:desc_version} presents such an example from the \textit{jackson-databind} project. The bug concerns certain fields being returned as null values even though they should have been skipped. The \textit{Bug description} explicitly states this failure condition, directly pointing the core issue. However, the version section only lists the software versions where the issue happens. If the model allocates substantial attention to this metadata, it may focus on strings of numbers that provide little guidance for diagnosing the bug while ignoring other relevant facts. The ground-truth patch, in this bug, simply adds a missing \texttt{`if (value == null)'} check. However, \textit{claude-4} fails to generate this condition, showing how emphasizing less informative metadata over the diagnostic description can lead to an incorrect repair.}

Another notable observation in our analysis is that \textit{Bug description} and \textit{Version information} are relatively consistent in structure across bug reports. The description typically contains natural language explanations of the failure, while version sections contain environment details. 
Other sections are far more heterogeneous in terms of content. For example, \textit{Reproduction} or \textit{Actual Behavior} may contain code snippets, stacktraces, or natural language explanations, which vary substantially across reports. 
Since the information in these sections differs substantially from one report to another, their section-level attention signals are likely diluted, resulting in smaller and less consistent effect sizes.
Based on these observations, it is possible that attention differences may be driven by finer-grained components within the sections rather than the sections themselves. We examine this more closely in \textit{RQ2}.

\begin{tcolorbox}[colback=gray!15, breakable, colframe=gray!40, left=1pt, right=1pt, top=1pt, bottom=1pt]
Across models, successful repairs consistently allocate most attention to the \textit{Bug description}, whereas unsuccessful repairs allocate more attention to \textit{Version information}. Other sections show weaker or no consistent differences.
\end{tcolorbox}

\subsection{RQ2: How is model attention distributed across specific code and natural-language components within bug report sections?}

\textbf{\underline{Results. }}
Table~\ref{tab:attn_or_fisher} summarizes the \textit{RQ2} results by quantifying how different attention patterns over bug report components are associated with repair success, both overall and by programming languages (Python and Java). For each pattern, we report Fisher’s exact test results (\textit{$p$}-value) with Benjamini-Hochberg FDR correction (FDR \textit{$q$}), logistic regression odds ratios with confidence intervals (OR [CI]), and the number of bugs in which the pattern appears (Count). The \textit{Influence} column indicates the direction of the association: \textit{OR} $>$ 1 denotes a positive association with successful repairs, whereas \textit{OR} $<$ 1 denotes a negative association.

We observe strong differences in overall repair success depending on how attention is distributed across bug report components.
\textbf{\textit{Diffused attention}}, where the model allocates attention across multiple components, is strongly associated with successful repairs ($\mathrm{OR}=2.07$, $q=2.41\times10^{-14}$). In contrast, \textbf{\textit{Localized attention}}, where the model concentrates on a specific part of the report, is negatively associated with repair success ($\mathrm{OR}=0.40$, $q=9.60\times10^{-15}$). These results suggest that successful repairs typically require integrating multiple sources of information from the bug report rather than focusing narrowly on a single component. When \textit{\textbf{no-attention}} is observed, we find no association with repair success or failure. 

This pattern is consistent across both Python and Java. Diffused attention remains positively associated with success in Python ($\mathrm{OR}=1.66$, $q=3.01\times10^{-8}$) and Java ($\mathrm{OR}=1.86$, $q=1.00\times10^{-6}$), while localized attention remains negatively associated in both Python ($\mathrm{OR}=0.40$, $q=1.28\times10^{-10}$) and Java ($\mathrm{OR}=0.65$, $q=4.71\times10^{-4}$). Although the negative association for localized attention is weaker in Java, the overall direction remains the same, suggesting that the relationship between attention structure and repair outcome is stable across programming languages.

Among components, code-related information within bug reports plays an important role in successful repair. In particular, attention to \textit{Code: Stacktrace} shows a significant positive association with repair success ($\mathrm{OR}=1.41$, $q=1.57\times10^{-2}$). This result is consistent with the diagnostic role of stacktraces in identifying potential failure locations~\cite{soltani_significance_2020}. Similarly, attention to \textit{Code: Test} is positively associated with success ($\mathrm{OR}=1.16$, $q=2.79\times10^{-2}$), indicating that models benefit from examples that concretely demonstrate the failing behavior. However, the relatively small count for test code (Count $=5$) suggests that this result should be interpreted more cautiously than higher-count components such as stacktraces (Count $=54$). Other code components, such as \textit{Code: Classes and methods} definitions and API-related usage (\textit{Code: Imports and variables}), show positive but statistically non-significant associations with repair success. While these elements may provide useful implementation context, their influence appears less consistent across bugs when compared to stacktrace or test information.

Among natural language components, attention to the \textit{NL: Description} shows a significant positive association with successful repair ($\mathrm{OR}=1.42$, $q=2.79\times10^{-2}$), showing that models benefit from detailed problem descriptions. In contrast, attention to \textit{NL: Version information} is negatively associated with success ($\mathrm{OR}=0.61$, $q=3.05\times10^{-2}$), suggesting that overemphasizing only on environmental metadata could be less relevant to produce correct patches. This is consistent with our \textit{RQ1} results, where successful repairs allocate more attention to the \textit{Bug description} section and less to \textit{Version information}, both of which mainly consist of natural language components.
Other natural language components, such as \textit{expected behavior}, \textit{reproduction steps}, and \textit{actual behavior}, show positive trends but do not remain statistically significant after multiple testing correction. The \textit{additional information} section shows a weak negative trend, likely because it frequently contains external links or references that are inaccessible to the model.

\textcolor{black}{Table~\ref{tab:attn_or_difficulty_control} reports the logistic regression results after controlling for bug difficulty. Overall, our findings remain highly consistent after accounting for difficulty. Diffused attention continues to show a strong positive association with repair success ($\mathrm{OR}=2.00$), while localized attention remains strongly negatively associated ($\mathrm{OR}=0.38$). Attention to \textit{NL: Description} ($\mathrm{OR}=1.37$), \textit{Code: Stacktrace} ($\mathrm{OR}=1.40$), and \textit{Code: Test} ($\mathrm{OR}=1.15$) remains positively associated with successful repair, whereas attention to \textit{NL: Version information} continues to show a negative association ($\mathrm{OR}=0.63$). These effect sizes are nearly identical to those observed in Table~\ref{tab:attn_or_fisher} (e.g., Diffused attention 2.07 vs. 2.00), indicating that the identified patterns are robust even after accounting for difficulty. Although \textit{Medium} and \textit{Hard} bugs are associated with lower repair success than \textit{Easy} bugs, the consistency of the coefficients shows that bug difficulty alone does not explain the observed behaviors.}

\begin{table}[t]
\centering
\footnotesize
\caption{\textcolor{black}{Associations between attention patterns and repair success while controlling for bug difficulty (\textit{Easy}: Baseline). Odds Ratios (OR) are reported with Confidence Intervals (CI).}}
\vspace{-0.4cm}
\label{tab:attn_or_difficulty_control}
\begin{tabular}{l c r l}
\toprule
\textbf{Pattern} & \textbf{Influence} & \textbf{Coef.} & \textbf{OR [CI]} \\
\midrule

\multicolumn{4}{@{}l@{}}{\textbf{\underline{Attention Structure}}} \\
\addlinespace[2pt]

\rowcolor{pos}
\textbf{Diffused} & \up
& 0.692 & \textbf{2.00 [1.54, 2.56]} \\

\rowcolor{neg}
\textbf{Localized} & \down
& -0.962 & \textbf{0.38 [0.30, 0.49]} \\

\midrule

\multicolumn{4}{@{}l@{}}{\textbf{\underline{Focus Targets}}} \\
\addlinespace[2pt]

\rowcolor{pos}
\textbf{NL:Description} & \up
& 0.313 & 1.37 [1.05, 1.81] \\

NL:Expected behavior & \up
& 0.260 & 1.30 [1.05, 1.61] \\

NL:Reproduction & \up
& 0.202 & 1.22 [0.95, 1.57] \\

NL:Actual behavior & \up
& 0.094 & 1.10 [0.94, 1.27] \\

NL:Additional information & \down
& -0.023 & 0.98 [0.77, 1.24] \\

\rowcolor{neg}
\textbf{NL:Version information} & \down
& -0.468 & 0.63 [0.50, 0.79] \\

\cmidrule(r){1-1}

\rowcolor{pos}
\textbf{Code:Stacktrace} & \up
& 0.338 & 1.40 [1.08, 1.78] \\

Code:Classes and methods & \up
& 0.226 & 1.25 [0.93, 1.66] \\

\rowcolor{pos}
\textbf{Code:Test} & \up
& 0.139 & 1.15 [1.03, 1.29] \\

Code:Imports and variables & \up
& 0.146 & 1.16 [0.88, 1.51] \\

\midrule
\multicolumn{4}{@{}l@{}}{\textbf{\underline{Bug Difficulty (Control Variables)}}} \\
\addlinespace[2pt]

Difficulty:Hard & \down
& -0.037 & 0.96 [0.78, 1.19] \\

Difficulty:Medium & \down
& -0.426 & 0.65 [0.50, 0.85] \\

\bottomrule
\end{tabular}
\vspace{-0.5cm}
\end{table}

\begin{figure}[h]
    \centering
    \includegraphics[width=0.95\linewidth]{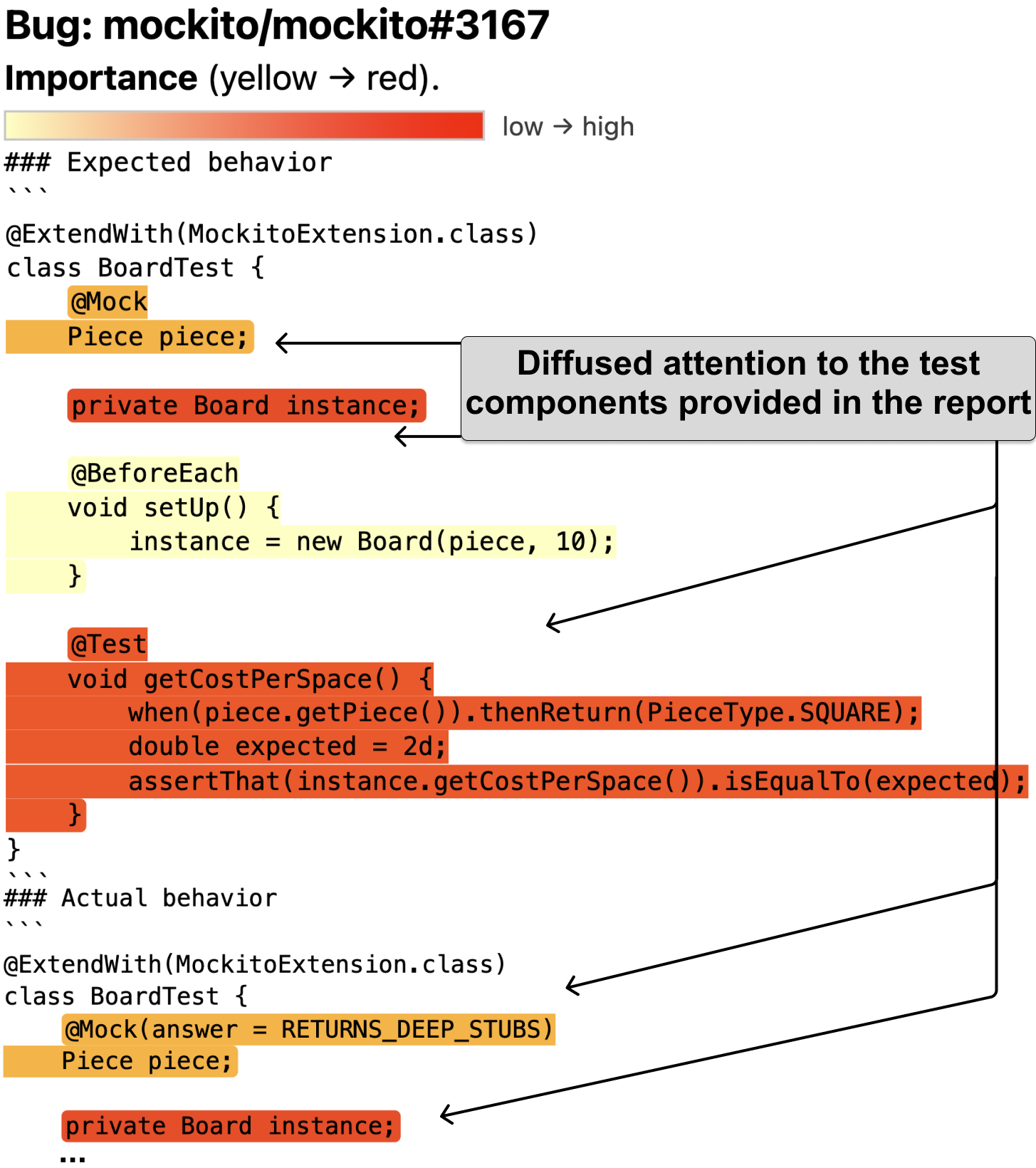}
    \vspace{-0.3cm}
    \caption{Example of Diffused Attention Throughout the Report in Project \textit{mockito}.}
    \label{fig:diff_tests}
\end{figure}

\noindent
\textbf{\underline{Discussion. }}\textcolor{black}{Diffused attention shows to be the most effective attention pattern for repair across bugs with different difficulties because the model integrates multiple components within the bug report instead of relying on a single source of information. Figure~\ref{fig:diff_tests} shows an example from project \textit{mockito}. In this case, the bug occurs from incorrect abstraction of enums, and the correct repair requires understanding both the failing test and the expected behavior. By distributing attention across the test information describing the current and expected behavior, the model is able to generate the correct patch, matching the ground-truth implementation. A similar pattern appears in Figure~\ref{fig:diffused_attention}, which shows a bug report from project \textit{xarray}. Here, the model attends to both natural language explanations and code components describing the intended logic. By combining these complementary signals, the model is able to correctly modify the function to accept more than one dimension, consistent with the ground-truth patch.}


\begin{figure}[h]
    \centering
    \includegraphics[width=0.95\linewidth]{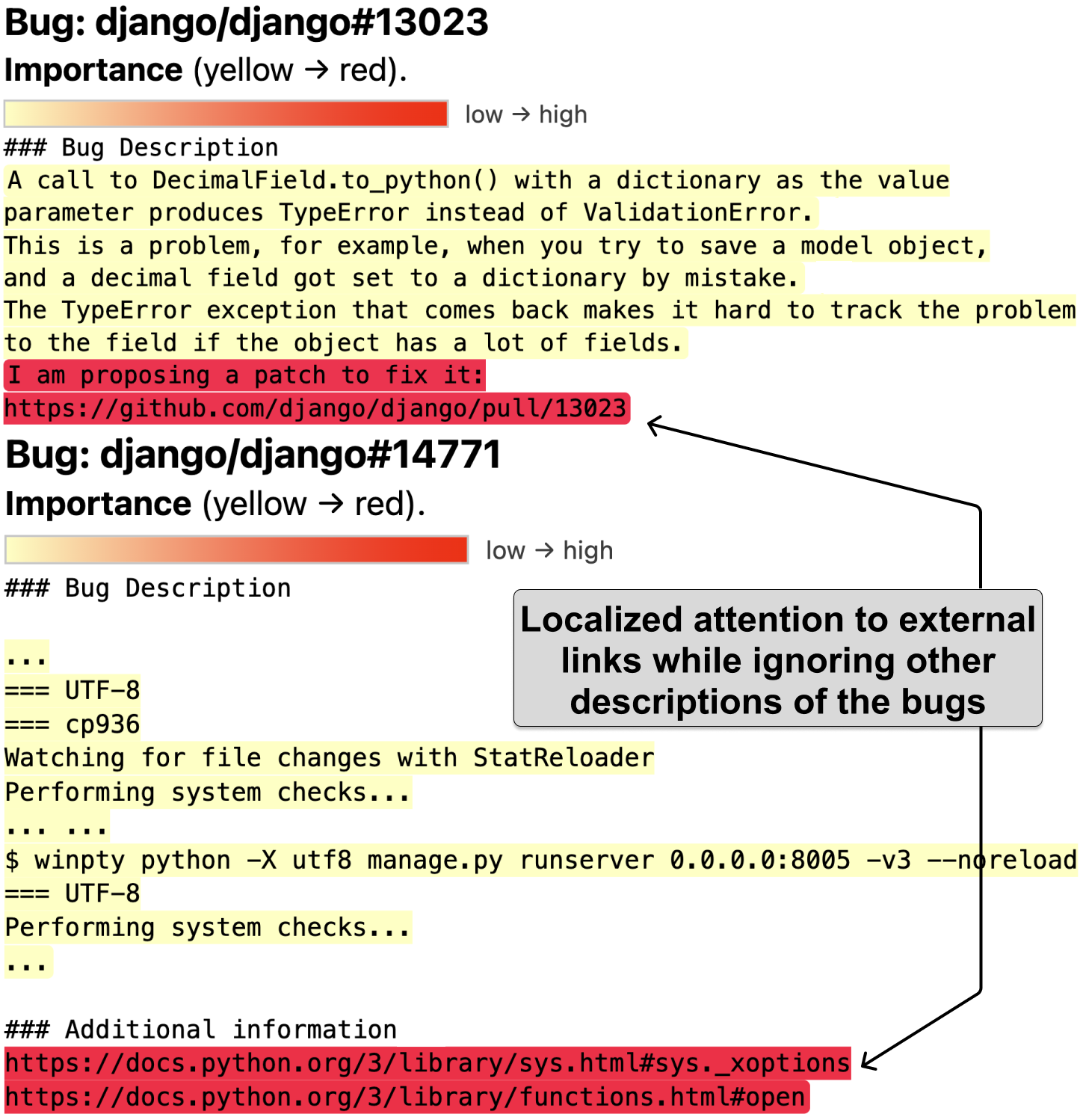}
\vspace{-0.35cm}
    \caption{Two Examples of Localized Attention in Reports of \textit{django} Project.}
    \label{fig:local_django}
\vspace{-0.5cm}
\end{figure}

\textcolor{black}{Localized attention, however, can act as a double-edged sword. Figure~\ref{fig:localized_attention} shows a bug from project \textit{xarray} where focusing on a specific developer-provided hint, the model generates the correct patch. However, this behavior often leads to failure when the localized components are misleading or incomplete. For example, Figure~\ref{fig:local_django} presents two bugs from the project \textit{django}. In the first case, the model fixates on a sentence referencing an external pull request that it cannot access while overlooking the earlier description explaining that the bug is caused by a \texttt{`TypeError'}. The ground-truth patch simply adds a missing \texttt{`TypeError'} check, but the generated patch misses this condition and therefore fails.
In the second case, the model concentrates on reference links in the additional information section while ignoring diagnostic system logs earlier in the report. The ground-truth patch updates the \texttt{`autoreloader'} to correctly pass the \texttt{`-X'} options described in these logs, whereas the generated patch fails to incorporate this behavior.
These examples show that while localized attention may occasionally succeed when the focused component directly encodes the solution, successful repairs more consistently emerge when models distribute attention across multiple complementary components within the bug report.}

\textcolor{black}{In addition, to assess whether these observations are specific to \textit{qwen3-32b}, we repeated our \textit{RQ2} analysis on the Java subset (71 bugs) using \textit{gpt-oss-20b}. As shown in Table~\ref{tab:gptoss20b_java_attn}, we observe the same overall attention patterns. Diffused attention remains positively associated with repair success ($\mathrm{OR}=1.56$), while localized attention is negatively associated ($\mathrm{OR}=0.54$), and diagnostically relevant components such as stacktraces ($\mathrm{OR}=1.24$) and natural-language descriptions ($\mathrm{OR}=1.19$) again show positive associations with successful repair. These consistent trends across different LLMs suggest that the broader attention behaviors identified in our study are not unique to a single model.}

\begin{table}[h]
\centering
\footnotesize
\setlength{\tabcolsep}{2pt}
\caption{\textcolor{black}{Associations between attention patterns and repair success for \textit{gpt-oss-20b} on Java subset.  Bold* denotes statistical significance ($p < 0.05$).}}
\vspace{-0.2cm}
\label{tab:gptoss20b_java_attn}
\begin{tabular}{l c c c c}
\toprule
\textbf{Pattern} & \textbf{Influence} & \textbf{OR [CI]} & \textbf{$p$-value} & \textbf{Count} \\
\midrule

\multicolumn{5}{@{}l@{}}{\textbf{\underline{Attention Structure}}} \\
\addlinespace[2pt]

No-attention & \neutraldir
& 1.0 [1.0, 1.0] & $4.54{\times}10^{-1}$ & 8 \\

\rowcolor{pos}
\textbf{Diffused*} & \up
& \textbf{1.56 [1.26, 1.90]} & $\mathbf{1.34{\times}10^{-5}}$ & 35 \\

\rowcolor{neg}
\textbf{Localized*} & \down
& \textbf{0.54 [0.45, 0.66]} & $\mathbf{1.09{\times}10^{-7}}$ & 28 \\

\midrule

\multicolumn{5}{@{}l@{}}{\textbf{\underline{Focus Targets}}} \\
\addlinespace[2pt]

NL:Description & \up
& 1.19 [0.94, 1.51] & $9.16{\times}10^{-2}$ & 40 \\

NL:Expected behavior & \up
& 1.13 [0.96, 1.32] & $8.39{\times}10^{-2}$ & 6 \\

NL:Reproduction & \up
& 1.06 [0.88, 1.28] & $3.30{\times}10^{-1}$ & 10 \\

NL:Actual behavior & \down
& 0.95 [0.84, 1.00] & $1.00$ & 1 \\

NL:Additional information & \up
& 1.06 [0.92, 1.23] & $6.93{\times}10^{-1}$ & 7 \\

NL:Version information & \down
& 0.92 [0.78, 1.09] & $6.83{\times}10^{-1}$ & 6 \\

\cmidrule(r){1-1}

\rowcolor{pos}
\textbf{Code:Stacktrace*} & \up
& \textbf{1.24 [1.02, 1.49]} & $\mathbf{1.77{\times}10^{-2}}$ & 11 \\

Code:Test & \up
& 1.15 [0.96, 1.35] & $8.39{\times}10^{-2}$ & 6 \\

Code:Imports and variables & \up
& 1.07 [0.90, 1.27] & $7.22{\times}10^{-1}$ & 9 \\

Code:Classes and methods & \down
& 0.90 [0.72, 1.12] & $1.00$ & 21 \\

\bottomrule
\end{tabular}
\vspace{-0.25cm}
\end{table}

\begin{tcolorbox}[colback=gray!15, breakable, colframe=gray!40, left=1pt, right=1pt, top=1pt, bottom=1pt]
Successful repairs are characterized by \textit{diffused attention} across multiple components, with strong focus on diagnostic information such as bug descriptions and stacktraces. In contrast, failures often involve more \textit{localized attention}, especially on non-diagnostic metadata.
\end{tcolorbox}

\subsection{RQ3: How well do model attention patterns align with what developers consider important for bug-fixing?}

\textbf{\underline{Results. }}Table~\ref{tab:section-frequency} reports how frequently each bug report section appears among the top-2 most important sections identified by developers and by the LLM. Overall, both developers and the LLM most frequently prioritize the \textit{Bug description} section, appearing in 79\% and 61\% of bugs, respectively, with a substantial overlap of 54\%. The \textit{Reproduction} section is the second most commonly prioritized (43\% for developers and 34\% for the LLM), showing a high overlap (30\%).
Other sections are selected less frequently. Developers emphasize \textit{Expected behavior} more often (18\%) than the LLM (11\%), while the LLM slightly prioritizes \textit{Version information} and \textit{Actual behavior} more than developers.

\begin{table}[h]
\centering
\footnotesize
\setlength{\tabcolsep}{4pt}
\renewcommand{\arraystretch}{1.1}
\caption{Frequency and Overlap of the top-2 most important sections selected by developers vs. LLM.}
\vspace{-0.4cm}
\label{tab:section-frequency}
\begin{tabularx}{\columnwidth}{@{}L r r r@{}}
\toprule
\textbf{Section} & \textbf{Developer Top-2 (\%)} & \textbf{LLM Top-2 (\%)} & \textbf{Overlap (\%)} \\
\midrule

Bug description        & 79 (79\%) & 61 (61\%) & 54 (54\%) \\
Reproduction           & 43 (43\%) & 34 (34\%) & 30 (30\%) \\
Expected behavior      & 18 (18\%) & 11 (11\%) & 6 (6\%) \\
Additional information & 16 (16\%) & 15 (15\%) & 8 (8\%) \\
Version information    & 6 (6\%)   & 11 (11\%) & 5 (5\%) \\
Actual behavior        & 4 (4\%)   & 7 (7\%)   & 2 (2\%) \\

\bottomrule
\end{tabularx}
\vspace{-0.2cm}
\end{table}

Table~\ref{tab:alignment-metrics} reports the mean alignment scores of successful and unsuccessful repairs, along with the Mann-Whitney $U$ test $p$-values.
At the section level, successful repairs exhibit significantly stronger alignment with developer importance ratings. The average Spearman correlation between developer Likert ratings and model attention is significantly higher for successful repairs (${rho}$ = 0.60) compared to unsuccessful ones (${rho}$ = 0.34), yielding a statistically significant difference ($p = 0.036$). This suggests that when models prioritize bug report sections similarly to developers, they are more likely to generate correct patches.
Hit@2 and Hit@1, which capture the overlap between the sections that developers and model rated highest, show that successful repairs achieve a higher average score (1.17 vs. 0.93, 0.57 vs. 0.33, respectively), with Hit@1 being statistically significant ($p = 0.02$).

\begin{table}[h]
\centering
\footnotesize
\setlength{\tabcolsep}{4pt}
\caption{Alignment metrics for successful vs. unsuccessful Repairs. Bold* denotes statistical significance ($p < 0.05$) with the Mann-Whitney U test. Column $\Delta$ shows the mean difference (Success-Fail).}
\vspace{-0.3cm}
\label{tab:alignment-metrics}

\begin{tabularx}{\columnwidth}{@{}l X l l l l@{}}
\toprule
\textbf{Level} & \textbf{Alignment Metric} & \textbf{Success} & \textbf{Fail} & $\boldsymbol{\Delta}$ & \textbf{$p$-value} \\
\midrule

\textbf{Section} & \textbf{Spearman $\rho$*} & 0.60 & 0.34 & 0.26\up & \textbf{0.036} \\
              & Hit@2                     & 1.17 & 0.93 & 0.24\up & 0.105 \\
              & \textbf{Hit@1*}                     & 0.57 & 0.33 & 0.24\up & \textbf{0.028} \\

\midrule

\textbf{Phrase}  & \textbf{F1@20*}        & 0.19 & 0.08 & 0.11\up & \textbf{0.031} \\
               & \textbf{Precision@20*} & 0.42 & 0.18 & 0.24\up & \textbf{0.027} \\
               & \textbf{Recall@20*}    & 0.13 & 0.05 & 0.08\up & \textbf{0.048} \\
               \cmidrule(lr){2-6}
               & \textbf{F1@10*}        & 0.10 & 0.04 & 0.06\up & \textbf{0.019} \\
               & \textbf{Precision@10*} & 0.37 & 0.16 & 0.21\up & \textbf{0.013} \\
               & \textbf{Recall@10*}    & 0.06 & 0.03 & 0.03\up & \textbf{0.023} \\

\bottomrule
\end{tabularx}
\vspace{-0.3cm}
\end{table}

A similar trend appears at the phrase level as well. Successful repairs consistently achieve higher overlap between model-attended phrases and developer key phrases across all metrics. Successful repairs show higher F1@20 (0.19 vs.\ 0.08, $p = 0.031$), Precision@20 (0.42 vs.\ 0.18, $p = 0.027$), and Recall@20 (0.13 vs.\ 0.05, $p = 0.048$) (same trend for k=10). These results indicate that models producing correct repairs are more likely to focus on the specific phrases that developers identify as important for diagnosing the bug.

\noindent
\textbf{\underline{Discussion. }}
Prior work has shown that during bug fixing, developers rely heavily on specific bug report elements, such as bug description, test cases, and stack traces~\cite{10.1145/3106237.3106285, 10.1145/3338906.3338947, 10.1109/TSE.2010.63}. Our findings extend this perspective to LLM-based program repair; LLMs are more likely to succeed when model attention aligns with the sections and phrases that developers consider most important.

\begin{figure}[h]
    \centering
    \includegraphics[width=0.95\linewidth]{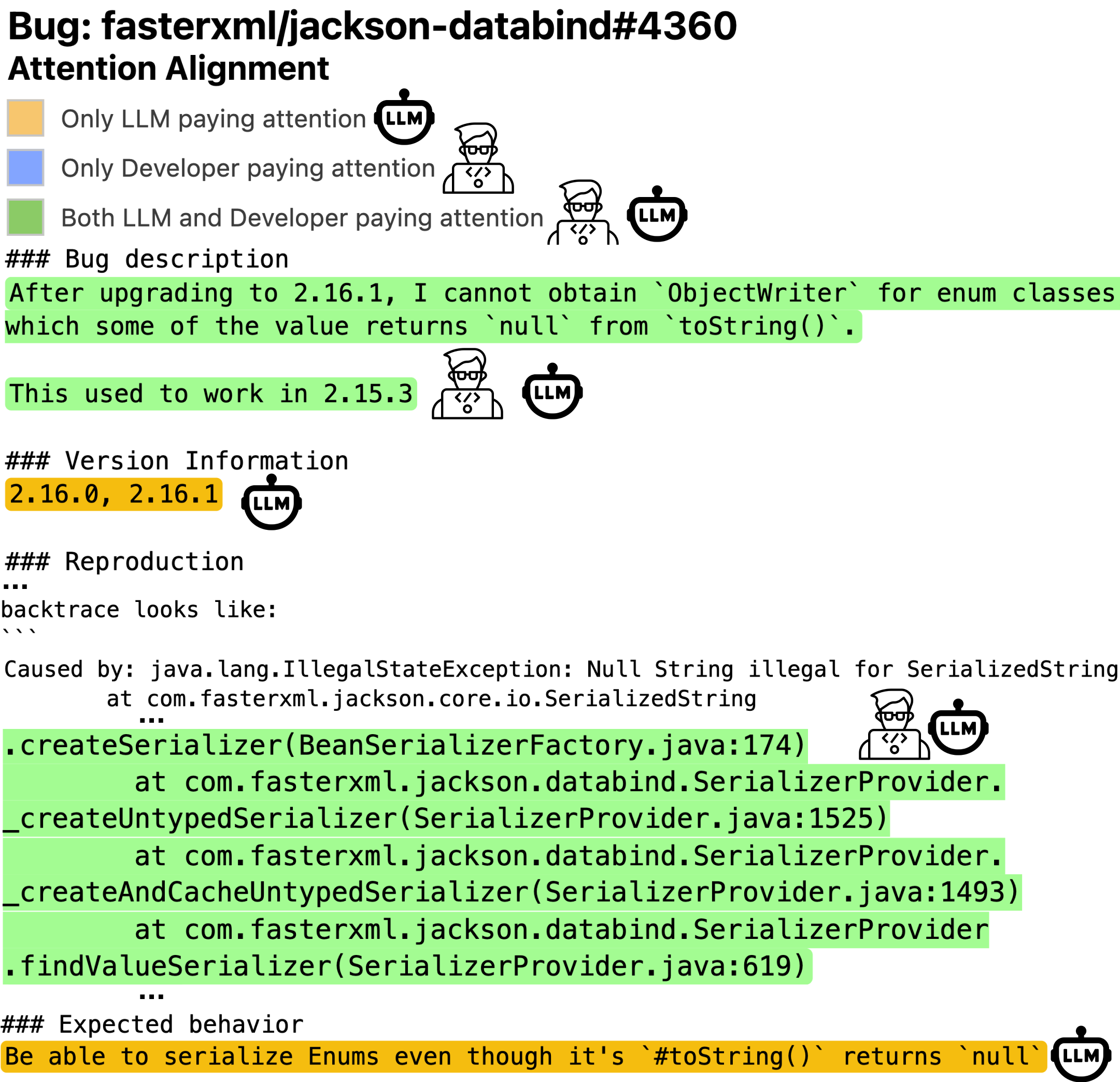}
    \vspace{-0.3cm}
    \caption{Aligned attention between developer and LLM for a bug in project \textit{jackson-databind}.}
    \label{fig:aligned_att}
    \vspace{-0.4cm}
\end{figure}

\begin{figure}[h]
    \centering
    \includegraphics[width=0.7\linewidth]{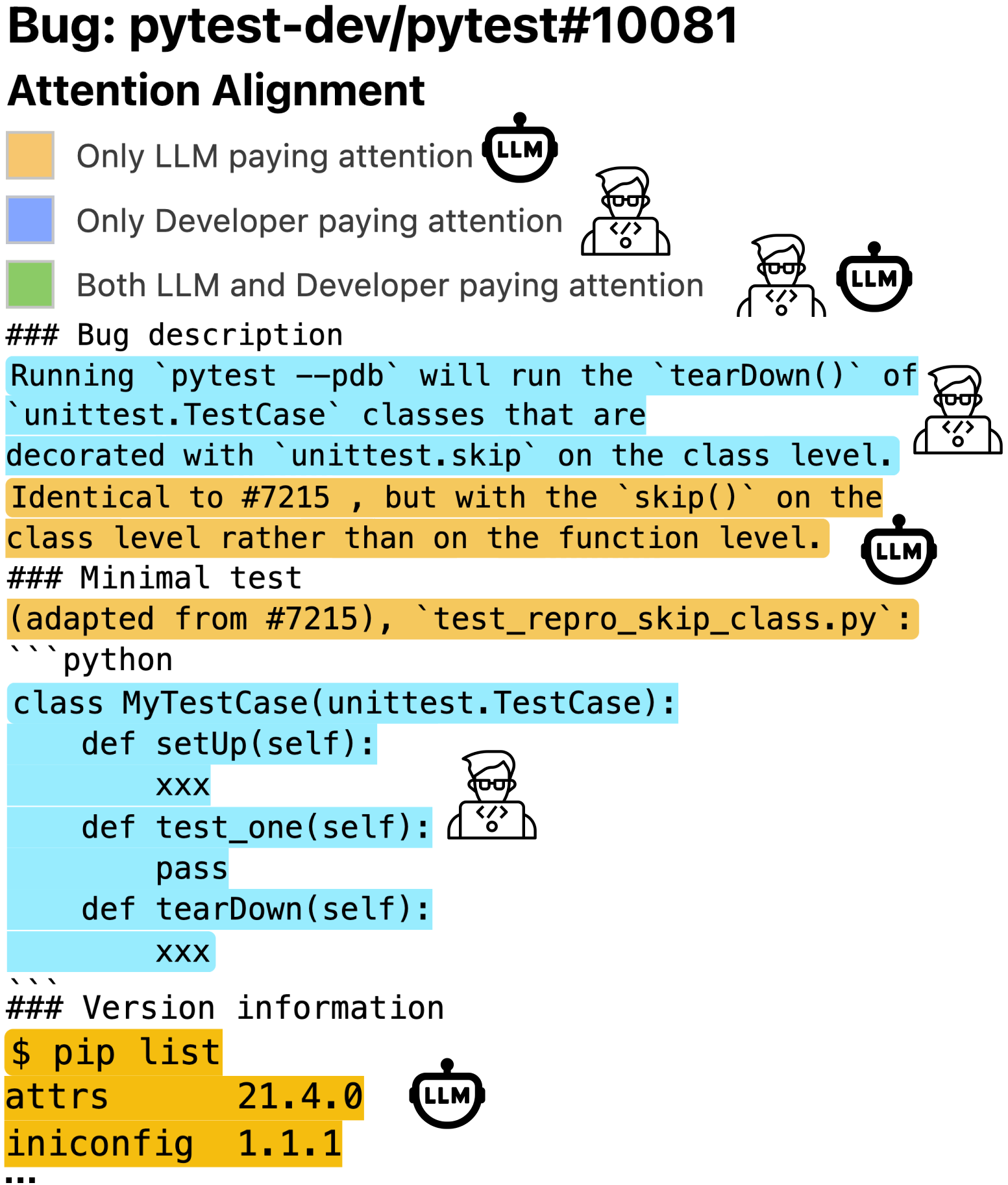}
    \vspace{-0.3cm}
    \caption{Misaligned attention between developer and LLM for a bug in project \textit{pytest}.}
    \label{fig:misaligned_att}
    \vspace{-0.3cm}

\end{figure}

\textcolor{black}{Figure~\ref{fig:aligned_att} shows an example of aligned attention in a bug report from project \textit{jackson-databind}. In this case, both the developer and model concentrate on the key descriptions of the bug and stacktrace. These components clearly reveal the underlying issue, which is the incorrect handling of enum serialization. And relying on this information, the model successfully generates a patch matching the ground-truth implementation.
However, Figure~\ref{fig:misaligned_att} shows an example from project \textit{pytest} where there is a clear attention divergence between model and developer. The developer focuses on the description and test information that reveal the failing behavior. The model pays attention to version information the most. While these components provide environmental context, they offer limited value for identifying the root cause of the bug. The ground-truth patch modifies the function so that tests are no longer executed when the class has been skipped using \texttt{`unittest.skip'} or \texttt{`pytest.mark.skip'}. However, the model fails to generate this behavior, resulting in an incorrect repair.}

Our finding adds an important task-specific perspective to prior work on human-model attention alignment~\cite{10645745, 10.1145/3691620.3695500}. Prior work in code generation has shown that human-model misalignment can help explain model errors~\cite{kou2024large, li2024machines, ning2024insights}. Our results extend this insight to program repair, suggesting that incorrect patches may arise when model attention diverges from the diagnostic evidence that developers consider important. Studying such alignment, therefore, makes repair behavior more interpretable. In future, our developer attention dataset could also serve as useful supervision for guiding the fine-tuning of APR models.

\begin{tcolorbox}[colback=gray!15, breakable, colframe=gray!40, left=1pt, right=1pt, top=1pt, bottom=1pt]
Stronger attention alignment between model and developers, both at the section and phrase levels, is associated with higher repair success.
\end{tcolorbox}

\section{Related Work}
\textbf{LLM-based Program Repair.}
Prior work has explored a wide range of techniques for improving the effectiveness of LLMs for APR. Early work showed that even static prompting with LLMs can outperform traditional APR tools~\cite{10.1109/ICSE48619.2023.00129}, and subsequent systems such as MMAPR~\cite{10.1145/3649850}, RING~\cite{10.1609/aaai.v37i4.25642}, and InferFix~\cite{10.1145/3611643.3613892} further improved repair performance using methods such as few-shot prompting.
Benchmarks built from real-world software bugs, such as \textit{SWE-bench}~\cite{jimenez2024swebench}, highlighted the challenges of applying LLMs to realistic bug repair tasks. However, studies showed that incorporating richer contextual signals into LLMs can significantly improve their performance. Fault localization signals~\cite{10.1109/ICSE48619.2023.00128}, failing tests~\cite{10.1145/3650212.3680323}, relevant code~\cite{10.1109/ICSE48619.2023.00125,10.1109/ASE56229.2023.00047}, and stacktraces~\cite{52980, haque-etal-2025-towards} can all improve repair accuracy. 
Historical information, such as prior commits, has also been shown to provide useful context for patch generation~\cite{shi2025hafixhistoryaugmentedlargelanguage}. 
Parasaram et al. proposed MANIPLE, which includes relevant bug-related facts from a repository into prompts to improve patch generation~\cite{10.1109/ICSE55347.2025.00162}. Ehsani et al. introduced a hierarchical knowledge injection technique that adds contextual information to LLMs in three layers (bug, repository, project) at a time, achieving significant improvements over MANIPLE~\cite{ehsani2025hierarchicalknowledgeinjectionimproving}.
More recently, agentic systems such as SWE-Agent~\cite{yang2024sweagent}, OpenHands~\cite{wang2025openhandsopenplatformai}, AutoCodeRover~\cite{10.1145/3650212.3680384}, and iSWE-agent~\cite{ganhotra2026resolvingjavacoderepository} combine multi-round reasoning, repository exploration, and tool interactions to iteratively generate and validate patches. 
At the time of writing, Sonar Foundation Agent~\cite{sonar} and live-SWE agent~\cite{xia2025livesweagentsoftwareengineeringagents} both achieve the highest repair performance on \textit{\swebenchverified{}}, resolving up to 79\% of bugs.

These works highlight the importance of contextual signals for LLM-based bug repair. However, it is unclear how LLMs interpret the information they receive.
To our knowledge, our work is the first to analyze how LLMs attend to different information in a bug report, and how these attention patterns relate to repair success.

\noindent
\textbf{Model Attention Analysis.}
Attention in transformer models and its role in interpretability have been extensively examined in previous work.
Early studies analyze attention weights across layers and heads to understand how semantic information is encoded~\cite{clark-etal-2019-bert, kovaleva-etal-2019-revealing}.
Other works propose alternative analysis techniques, including perturbation-based methods~\cite{serrano-smith-2019-attention}, gradient-based attribution~\cite{jain-wallace-2019-attention, bastings2020elephantinterpretabilityroomuse}, and attention aggregation strategies~\cite{clark-etal-2019-bert, kovaleva-etal-2019-revealing}, to characterize what attention represents and how it relates to model behavior.
More recent efforts extend these ideas to large language models to improve interpretability~\cite{zhou2025roleattentionheadslarge, zheng2024attentionheadslargelanguage, 11334352}, reduce hallucination~\cite{huang2025risk}, and improve inference efficiency through attention optimization~\cite{li2024snapkv}.
Recent work has also examined attention in LLM-based code generation. Kou et al.~\cite{kou2024large} show that models often attend to different parts of task descriptions than humans during code generation, highlighting the need for human-aligned LLMs for better interpretability and developer trust.  
Other works also observed misalignment between model attention and human reasoning, suggesting that attention patterns can reveal failure modes in generated code~\cite{li2024machines, 10645745, ning2024insights}.
Several studies explore methods to align model attention with human signals, such as using eye-tracking data for fine-tuning~\cite{zhang2025eyemulator} or structure-aware attention mechanisms to guide models toward meaningful input regions~\cite{liang-etal-2025-waffle}. Overall, these works suggest that attention alignment can not only improve interpretability, but also lead to better task performance~\cite{10645745, 10.1145/3691620.3695500}.

Despite these advances, interpretability remains a major challenge for integrating LLMs reliably into developer workflows~\cite{10.1145/3704806, li2025riseaiteammatessoftware}. We aim to address this gap by analyzing how LLMs allocate attention over bug reports during repair, and how that compares to information developers consider important for bug fixing.
\section{Threats to Validity}

\textbf{Construct Validity.}
We measure model attention using perturbation analysis by removing parts of the bug report and observing how the model's output changes. While this provides a causal signal (changes in input lead to changes in output), it is still an approximation and may not fully reflect model reasoning. However, prior work shows that perturbation-based methods provide a more direct and human-aligned interpretation compared to other attention analysis methods~\cite{kou2024large,jain-wallace-2019-attention, 11334352}.
\textcolor{black}{For \textit{RQ3}, we use developer judgments to identify important sections and phrases in bug reports. 
To reduce subjectivity in these manual annotations, we developed annotation guidelines through multiple discussions, recruited experienced developers, conducted a pilot study, and assigned an equal number of bugs to each annotator. We also computed inter-rater agreement on a subset of annotations, showing strong agreement (Spearman $=0.81$, Cohen's Kappa $=0.77$).}

\noindent
\textbf{Internal Validity.}
We use deterministic decoding (zero temperature) and fixed prompts to reduce randomness and isolate the effect of input perturbations.
Although attention may vary under stochastic decoding, this controlled setting supports more reliable analysis. Because perturbation removes parts of a bug report, masking may disrupt information flow; we mitigate this by applying the same masking procedure across distinct sections.
\textcolor{black}{Repair outcomes may also be influenced by factors such as bug difficulty. To mitigate this threat, in \textit{RQ2}, we repeat our analysis while controlling for bug difficulty, which results in highly identical results.}
In \textit{RQ3}, each bug report is annotated by a single developer because detailed manual annotation is costly and our goal is to capture subjective notions of importance that can naturally vary across developers.

\noindent
\textbf{External Validity.}
\textcolor{black}{Our dataset includes 319 Python and Java bugs from \textit{SWE-Bench Verified} and \textit{Multi-SWE-Bench}. While our findings might not generalize to other languages, domains, or less-structured bug reports, these widely used APR benchmarks consist of real-world software bugs and provide sufficiently detailed reports for controlled attention analysis.}
In \textit{RQ1}, we evaluate three LLMs (one proprietary and two open-source), and for a deeper analysis in \textit{RQ2} and \textit{RQ3}, we focus on one LLM due to the substantially higher computational cost of perturbation-based analysis. While these models represent different sizes and types, our findings may not generalize to other LLMs or agentic APR systems. \textcolor{black}{To mitigate this threat, we first compare section-level attention across all three models in \textit{RQ1}, observing consistent trends despite differences in repair performance. We further validate the generality of the fine-grained analysis by repeating \textit{RQ2} on a Java subset (71 bugs) using \textit{gpt-oss-20b}, obtaining the same overall attention patterns as \textit{qwen3-32b}.}
In \textit{RQ3}, our analysis focuses on a subset of 100 manually annotated bug reports, which might not generalize to larger datasets. However, for our dataset, this sample size is statistically significant, providing approximately 95\% confidence with a margin of error of $\pm$8\%~\cite{cochran1977sampling}.
\section{Conclusion and Future Work}
Using perturbation-based analysis over 319 real-world bugs, we examined how models allocate attention across bug reports, how these patterns differ between successful and unsuccessful repairs, and how model attention aligns with what information developers consider important for bug fixing.
Our findings show that repair success is strongly associated with how models prioritize information within bug reports. Successful repairs exhibit diffused attention over \textcolor{black}{multiple fine-grained information in} diagnostically important components, while failures often arise from over-localized attention toward less informative components. Stronger alignment between model attention and developers is linked to higher repair success, further highlighting attention misallocation as a key factor for failures in LLM-based program repair. \textcolor{black}{While bug difficulty and other factors could influence repair performance, our results show that they do not fully explain repair outcomes. Even after controlling for bug difficulty, the same attention patterns remain strongly associated with successful repairs, which shows the importance of understanding how LLMs process information during repair.}

\textcolor{black}{By revealing which parts of bug reports models prioritize during repair, our approach can help practitioners redesign prompts, retrieval strategies, and pre-processing pipelines to emphasize diagnostically useful information while reordering or reformulating components that consistently distract models or get ignored. Beyond prompt design, these insights can support attention-aware context selection and monitoring mechanisms that identify potential attention misallocation before incorrect repairs are generated. Our annotated dataset can also support future fine-tuning approaches that align LLM information prioritization with developer reasoning. Finally, because our perturbation-based method is model-agnostic and does not require access to internal model states, it can be applied to agentic APR systems to improve their interpretability and better understand how information is processed through complex repair workflows.}

\section*{Data Availability Statement}
Our replication package is available online at this DOI link: \url{https://doi.org/10.5281/zenodo.21381449}

\balance
\bibliographystyle{ACM-Reference-Format}
\bibliography{ref, apr}


\end{document}